\documentclass[12pt, notitlepage]{article}

\usepackage{a4}

\usepackage{natbib}

\usepackage{xcolor}

\definecolor{TUMblue}{RGB}{0, 101, 189}
\definecolor{TUMlightblue}{RGB}{100,160,200}
\definecolor{TUMdarkblue}{RGB}{0,51,89}
\definecolor{TUMgreen}{RGB}{162,173,0}
\definecolor{TUMorange}{RGB}{227,114,034}
\definecolor{TUMivory}{RGB}{218,215,203}

\usepackage{hyperref}

\hypersetup{
	colorlinks=true,
	linkcolor=TUMblue,
	citecolor=TUMblue,
	filecolor=TUMblue,
	urlcolor=TUMblue
}

\usepackage{etoolbox}
\makeatletter

\pretocmd{\NAT@citex}{%
	\let\NAT@hyper@\NAT@hyper@citex
	\def\NAT@postnote{#2}%
	\setcounter{NAT@total@cites}{0}%
	\setcounter{NAT@count@cites}{0}%
	\forcsvlist{\stepcounter{NAT@total@cites}\@gobble}{#3}}{}{}
\newcounter{NAT@total@cites}
\newcounter{NAT@count@cites}
\def\NAT@postnote{}

\def\NAT@hyper@citex#1{%
	\stepcounter{NAT@count@cites}%
	\hyper@natlinkstart{\@citeb\@extra@b@citeb}#1%
	\ifnumequal{\value{NAT@count@cites}}{\value{NAT@total@cites}}
	{\ifNAT@swa\else\if*\NAT@postnote*\else%
		\NAT@cmt\NAT@postnote\global\def\NAT@postnote{}\fi\fi}{}%
	\ifNAT@swa\else\if\relax\NAT@date\relax
	\else\NAT@@close\global\let\NAT@nm\@empty\fi\fi
	\hyper@natlinkend}
\renewcommand\hyper@natlinkbreak[2]{#1}

\patchcmd{\NAT@citex}
{\ifNAT@swa\else\if*#2*\else\NAT@cmt#2\fi
	\if\relax\NAT@date\relax\else\NAT@@close\fi\fi}{}{}{}
\patchcmd{\NAT@citex}
{\if\relax\NAT@date\relax\NAT@def@citea\else\NAT@def@citea@close\fi}
{\if\relax\NAT@date\relax\NAT@def@citea\else\NAT@def@citea@space\fi}{}{}

\makeatother

\usepackage{enumitem}
\usepackage{float}
\usepackage{tikz}
\usetikzlibrary{calc,shapes,arrows,decorations.pathmorphing,graphs,positioning,backgrounds}

\usepackage{tabulary}
\newcolumntype{K}[1]{>{\centering\arraybackslash}p{#1}}
\newcommand\Tstrut{\rule{0pt}{2.6ex}}         
\newcommand\Bstrut{\rule[-0.9ex]{0pt}{0pt}}   

\usepackage{amssymb, graphicx}
\usepackage{subfigure}
\usepackage{amsmath}
\usepackage{verbatim}
\usepackage{bbm}
\usepackage{ dsfont }
\usepackage{geometry}
\usepackage{amsthm}
\usepackage{aliascnt}
\newcommand{\mynewtheorem}[2]{
	\newaliascnt{#1}{dummy}
	\newtheorem{#1}[#1]{#2}
	\aliascntresetthe{#1}

	\expandafter\def\csname #1autorefname\endcsname{#2}
}

\newtheoremstyle{mystyle}
{}
{}
{}
{}
{\bfseries\sffamily}
{.}
{ }
{}

\theoremstyle{mystyle}
\mynewtheorem{thm}{Theorem}
\mynewtheorem{defi}{Definition}
\mynewtheorem{lem}{Lemma}
\mynewtheorem{cor}{Corollary}
\mynewtheorem{prop}{Proposition}
\mynewtheorem{exa}{Example}
\mynewtheorem{alg}{Algorithm}
\mynewtheorem{rem}{Remark}
\mynewtheorem{bsp}{Example}

\def\equationautorefname~#1\null{Equation~(#1)\null}
\newcommand{\appref}[1]{\hyperref[#1]{Appendix~\ref{#1}}}

\usepackage{multirow}
\usepackage{bigdelim}

\usepackage{sectsty}
\allsectionsfont{\sffamily}

\usepackage{url}

\usepackage{caption}
\newcommand{\Ebb}{\mathbb{E}}

\newcommand{\Nbb}{\mathbb{N}}

\newcommand{\Rbb}{\mathbb{R}}

\newcommand{\dint}{\mathrm{d}}

\newcommand{\betab}{\boldsymbol{\beta}}

\newcommand{\gammab}{\mathbf{\boldsymbol{\gamma}}}

\newcommand{\thetab}{\boldsymbol{\theta}}
\newcommand{\ub}{\mathbf{u}}
\newcommand{\Ub}{\mathbf{U}}

\newcommand{\xb}{\mathbf{x}}

\newcommand{\Yb}{\mathbf{Y}}
\newcommand{\yb}{\mathbf{y}}

\newcommand{\zb}{\mathbf{z}}
\newcommand{\0}{\mathbf{0}}
\newcommand{\1}{\mathbf{1}}

\newcommand{\Cc}{\mathcal{C}}

\newcommand{\Nc}{\mathcal{N}}

\newcommand{\Uc}{\mathcal{U}}

\newcommand{\Yc}{\mathcal{Y}}

\newcommand{\be}{\begin{equation}}
\newcommand{\ee}{\end{equation}}

\newcommand{\eps}{\varepsilon}
\newcommand{\epsb}{\boldsymbol{\varepsilon}}

\DeclareMathOperator{\argmax}{arg\,max}

\DeclareMathOperator{\BIC}{BIC}

\DeclareMathOperator{\diag}{diag}

\DeclareMathOperator{\Corr}{Cor}

\DeclareMathOperator{\Var}{Var}

\newcommand{\BIGOP}[1]{\mathop{\mathchoice%
		{\raise-0.22em\hbox{\huge $#1$}}%
		{\raise-0.05em\hbox{\Large $#1$}}{\hbox{\large $#1$}}{#1}}}

\begin{document}

\title{\textbf{\sffamily 
		A D-vine copula based model for repeated measurements extending linear mixed models with homogeneous correlation structure
		\vspace{2cm}}}

\date{\small \today}
\author{Matthias Killiches\footnote{Zentrum Mathematik, Technische Universit\"at M\"unchen, Boltzmannstra\ss e 3, 85748 Garching, Germany} \footnote{Corresponding author, email: \url{matthias.killiches@tum.de}.} \and Claudia Czado$^*$}

\maketitle
\vspace*{-5mm}
\begin{abstract}
	We propose a model for unbalanced longitudinal data, where the univariate margins can be selected arbitrarily and the dependence structure is described with the help of a D-vine copula. We show that our approach is an extremely flexible extension of the widely used linear mixed model if the correlation is homogeneous over the considered individuals. As an alternative to joint maximum-likelihood a sequential estimation approach for the D-vine copula is provided and validated in a simulation study. The model can handle missing values without being forced to discard data. Since conditional distributions are known analytically, we easily make predictions for future events. For model selection we adjust the Bayesian information criterion to our situation. In an application to heart surgery data our model performs clearly better than competing linear mixed models.\\
	
\end{abstract}

\section{Introduction}\label{sec:intro}
Repeated measurements that are obtained in a longitudinal study are common in many areas. Very early applications in astronomy \citep{airy1861algebraic} were followed by a vast number of studies in fields such as industry \citep[e.g.][]{newbold1927practical}, ecology \citep[e.g.][]{potvin1990statistical}, biology \citep[e.g.][]{yeung2003clustering}, psychology \citep[e.g.][]{lorch1990regression}, medicine \citep[e.g.][]{ludbrook1994repeated}, education \citep[e.g.][]{malin2001multilevel} and many more.

Over the years many concepts have been developed for the analysis of such repeated measurements. An extensive review on the origins of longitudinal data models can be found in Chapter 1 of \cite{fitzmaurice2008longitudinal}. \cite{davis2002statistical} offers a thorough introduction to the topic, starting with basic aspects of repeated measurement data. Besides foundations and different modeling aspects of repeated measurement data \cite{lindsey1999models} addresses the question how to design a study. \cite{diggle1989selected} give an extensive review on different approaches to the analysis of repeated measurements. The most popular model class for this purpose are probably \emph{linear mixed models} (LMMs). They extend classical linear models by adding individual-specific random effects to the fixed effects. Extensive introductions to this topic can be found for example in \cite{diggle2002analysis} and \cite{verbeke2009linear}. Although the covariance structure of linear mixed models can be fitted rather flexibly, the dependence always remains Gaussian by definition.

Within the last two decades dependence modeling has become more and more popular in all areas of applications. Especially copulas have gained large popularity since they allow to model marginal distributions and the dependence structure separately \citep{Sklar}. Consequently, copulas were also applied for modeling repeated measurement data. This approach has first been used by \cite{meester1994fully} who developed a model for bivariate clustered categorical data. \cite{lambert2002copula} present a model for multivariate repeated measurement data, where the dependence is described by copula (although only the Gaussian copula is used in the application). \cite{shen2006copula} model serial dependence for continuous longitudinal data with a non-ignorable non-monotone missing-data process using a Gaussian copula. Other examples are \cite{lindsey2006multivariate}, who use the Gaussian copula among other multivariate models with correlation matrices for non-linear repeated measurements. Further, \cite{sun2008heavy} argue that elliptical copulas are better suited than Archimedean copulas for modeling serial dependence in the context of longitudinal data.

\emph{D-vine copulas} are a special class of \emph{vine copulas} \citep{bedford2002vines,aasczado} that are particularly suited for modeling serial dependence. \cite{smith2010modeling} used them to model longitudinal data in a Bayesian approach. Multivariate time series are considered in \cite{smith2015copula} and \cite{NaiRuscone2017}. In \citet[][Chapter 7.5]{joe2014dependence} discrete longitudinal count data are modeled using D-vines. \cite{shi2016pair} use a mixed D-vine to model semi-continuous longitudinal claims. All these references work in a balanced setting, i.e.\ each individual has the same number of measurements. An unbalanced setting is considered by \cite{shi2016multilevel} using a Gaussian copula.

The novelty of the approach presented in this paper is that we develop a D-vine copula based model with arbitrary margins for modeling unbalanced longitudinal data with the aim of understanding the underlying relationship among the measurements and enabling predictions for future events. For prediction we use conditional quantiles that are analytically given. For model selection we derive an adjustment of the \emph{Bayesian information criterion} (BIC) for the proposed model. The model will furthermore be shown to be an extension of a very wide class of linear mixed models for which the correlation matrix of the measurements is homogeneous over the individuals. 

\autoref{sec:modeling} briefly introduces D-vine copulas and the proposed D-vine copula based model for repeated measurement data. Linear mixed models and their connection with the D-vine based model are developed in \autoref{sec:connectionLMM}. \autoref{sec:estimation} contains maximum-likelihood based estimation methods for the D-vine based model. Further, as a tool for model selection, an adjustment of the BIC for the proposed model is derived. The performance of the estimation methods is investigated in a simulation study (\autoref{sec:simstudy}). In \autoref{sec:application} we fit both linear mixed models and D-vine based models to a heart surgery data set and compare the results using likelihood based model selection criteria and performing conditional quantile prediction. \autoref{sec:conclusion} contains our conclusions and an outlook on future research.

\section{D-vine based repeated measurement model}\label{sec:modeling}

\subsection{Setting and marginal modeling}\label{sec:setting}

Consider a repeated measurement (longitudinal) data set $\Yc=\left\{\yb^1,\ldots,\yb^n\right\}$ that contains $n\in\Nbb$ observation blocks $\yb^i=(y^i_1,\ldots,y^i_{d_i})^\top\in\Rbb^{d_i}$ associated with individual $i$ having $d_i\in\left\{1,\ldots,d\right\}$ measurements. Here $d\in \Nbb$ denotes the maximum number of measurements per individual observed. For two different individuals the $j$th event does not necessarily need to have occurred at the same time $t_j$. We denote by $n_j$ the number of observations of length $j$, $j=1,\ldots,d$, where $n_j$ is zero if $\Yc$ contains no observations of length $j$. 
We divide now the data set into subsets of groups of individuals with the same number of measurements. For $j=1,\ldots,d$, we summarize the observations of group $j$ as $\Yc^j=\left\{\yb^i\mid i\in I_j\right\}$, where the corresponding index set is defined as $I_j=\left\{i\mid \yb^i\in\Rbb^{j}\right\}$. \autoref{tab:RE4d} illustrates the above notation and data structure for an exemplary data set of size $n=9$, where the maximum number of measurements per individual is $d=4$ and we have $n_1=0$ individuals with 1 measurement, $n_2=3$ individuals with 2 measurements, $n_3=2$ individuals with 3 measurements and $n_4=4$ individuals with 4 measurements. Consequently, $I_1=\emptyset$, $I_2=\left\{1,2,3\right\}$, $I_3=\left\{4,5\right\}$ and $I_4=\left\{6,7,8,9\right\}$.\\

\begin{table}[h!]
	\centering
	\begin{tabular}{cc|K{1.2cm}K{1.2cm}K{1.2cm}K{1.2cm}}
		\multicolumn{2}{c|}{observations} & \multicolumn{4}{c}{measurements}\\
		& & 1 & 2 & 3 & 4 \Bstrut\\
		\hline 
		\ldelim\{{3}{93pt}[$\Yc^2=\left\{\yb^i\mid i\in I_2\right\}\,\,$] & $\yb^1$ & $*$ & $*$ & &  \Tstrut \\
		& $\yb^2$ & $*$ & $*$ & & \\
		& $\yb^3$ & $*$ & $*$ & &  \\[2mm]
		\ldelim\{{2}{93pt}[$\Yc^3=\left\{\yb^i\mid i\in I_3\right\}\,\,$] & $\yb^4$ & $*$ & $*$ & $*$ & \\
		& $\yb^5$ & $*$ & $*$ & $*$ & \\[2mm]
		\ldelim\{{4}{95pt}[$\Yc^4=\left\{\yb^i\mid i\in I_4\right\}\,\,$] & $\yb^6$ & $*$ & $*$ & $*$ & $*$ \\
		& $\yb^7$ & $*$ & $*$ & $*$ & $*$ \\
		& $\yb^8$ & $*$ & $*$ & $*$ & $*$ \\
		& $\yb^9$ & $*$ & $*$ & $*$ & $*$ \\
	\end{tabular}
	\caption{Grouping of an exemplary data set of size $n=9$ with $d=4$, $n_2=3$, $n_3=2$ and $n_4=4$. Stars indicate observed events.}
	\label{tab:RE4d}
\end{table}

Having Sklar's Theorem \citep{Sklar} in mind, we follow a two-stage approach, also referred to as the Inference Functions for Margins (IFM) method \citep[cf.][Section 10.1]{joe1997multivariate}: First we use the probability integral transform and apply the univariate marginal distributions $F^i_j$ to the measurements $y_j^i\in\Rbb$ in order to transform them to measurements $u_j^i:=F^i_j(y_j^i)\in[0,1]$ to the uniform scale, $j=1,\ldots,d_i$ and $i=1,\ldots,n$. Then we model the dependence structure of the resulting uniform scale data utilizing a copula. In the following sections we will use a notation for the copula data that is similar to the one for the original data. The copula data $\Uc=\left\{\ub^1,\ldots,\ub^n\right\}$ consists of the observations $\ub^i=(u_1^i,\ldots,u_{d_i}^i)^\top\in[0,1]^{d_i}$, $i=1,\ldots,n$. Again, we form groups $\Uc^j=\left\{\ub^i\mid i\in I_j\right\}$ containing all observations of length $j$, $j=1,\ldots,d$. Since individuals with only one measurement do not contribute to the dependence structure we will only consider $\Uc^2,\ldots,\Uc^d$. Thus we can assume that $n_1=0$, i.e.\ $\Uc^1=\emptyset$, without losing generality. Of course, in practice the distribution functions $F^i_j$ are usually not known and need to be estimated (see \autoref{sec:estimation}).

\subsection{D-vine based dependence model}\label{sec:dependencemodel}
\paragraph{D-vine copulas}\label{sec:dvine}\mbox{}\\
Since we will use D-vine copulas for modeling the dependence we first give a short introduction to this model class. Vines have been introduced by \cite{bedford2002vines}. They are graphical models that can be used to construct a multivariate copula density as a product over bivariate building blocks, so-called pair-copulas. Since \cite{aasczado} presented statistical inference methods for vine copulas, their popularity has increased drastically. D-vines are a subclass of vines representing a sequential structure. They are frequently used \citep[e.g.][]{ren2014sequential, kim2013mixture, czado2011analysis} because of their flexibility and interpretability.

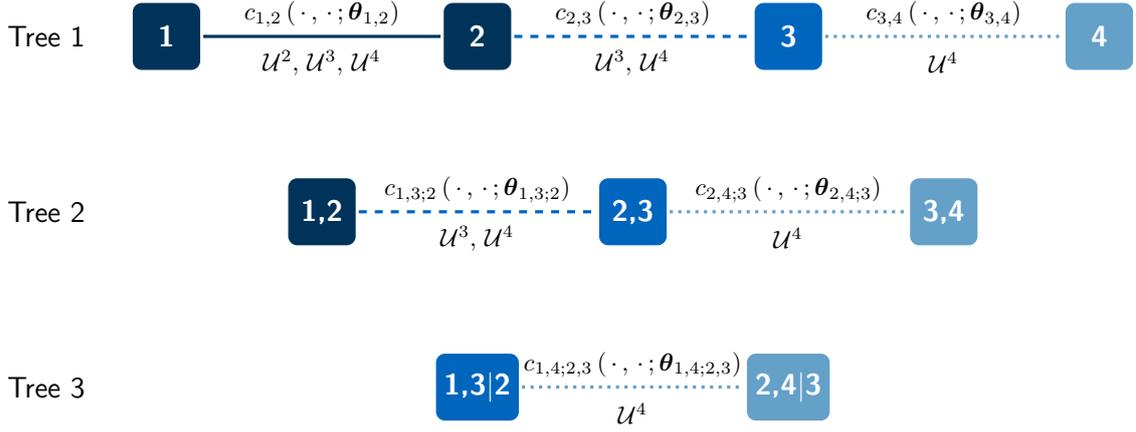
\begin{figure}
	\tikzstyle{VineNode} = [rounded corners, fill = white, draw = white, text = white, font = \bfseries\sffamily, align = center, minimum height = 1cm, minimum width = 1cm, very thick]
	\tikzstyle{TreeLabels} = [draw = none, fill = none, text = black]
	\tikzstyle{DummyNode}  = [draw = none, fill = none, text = white]
	\newcommand{\yshift}{-.3cm}
	\newcommand{\yshiftLabel}{-0.2cm}
	\newcommand{\labelsize}{\normalsize}
	\newcommand{\fontsizeEdge}{\footnotesize}
	\center
	\begin{tikzpicture}	[scale=0.93, every node/.style = {VineNode, scale=0.93}, node distance =2.2cm]
	
	\node (1)[fill = TUMdarkblue]{1}
	node[DummyNode] (D1-2) [right of = 1]{}
	node  (2)   [right of = D1-2, fill = TUMdarkblue]{2}
	node[DummyNode] (D2-3) [right of = 2]{}
	node  (3)   [right of = D2-3, fill = TUMblue]{3}
	node[DummyNode] (D3-4) [right of = 3]{}
	node  (4)   [right of = D3-4, fill = TUMlightblue]{4}
	
	node (1-2)  [below of = D1-2, yshift = \yshift, fill = TUMdarkblue]{1,2}
	node[DummyNode] (D1-3_2) [below of = 2, yshift = \yshift]{}
	node  (2-3)   [right of = D1-3_2, fill = TUMblue]{2,3}
	node[DummyNode] (D2-4_3) [right of = 2-3]{}
	node  (3-4)   [below of = D3-4, yshift = \yshift, fill = TUMlightblue]{3,4}
	
	node (1-3_2) [below of = D1-3_2, yshift = \yshift, fill = TUMblue]{1,3$|$2}
	node[DummyNode] (D1-4_2-3) [below of = 2, yshift = \yshift]{}
	node (2-4_3) [below of = D2-4_3, yshift = \yshift, fill = TUMlightblue]{2,4$|$3}
	;
	\draw[very thick, TUMdarkblue] (1) to node[draw=none, fill = none, font = \fontsizeEdge,
	below, yshift = 0.15cm, text = black] {$\Uc^2$, $\Uc^3$, $\Uc^4$}
	node[draw=none, fill = none, font = \fontsizeEdge,
	above, yshift = \yshiftLabel, text = black] {$c_{1,2}\left(\,\cdot\,,\,\cdot\,;\thetab_{1,2}\right)$} (2);
	\draw[dashed, very thick, TUMblue] (2) to node[draw=none, fill = none, font = \fontsizeEdge,
	below, yshift = 0.15cm, text = black] {$\Uc^3$, $\Uc^4$}
	node[draw=none, fill = none, font = \fontsizeEdge,
	above, yshift = \yshiftLabel, text = black] {$c_{2,3}\left(\,\cdot\,,\,\cdot\,;\thetab_{2,3}\right)$} (3);
	\draw[dotted, very thick, TUMlightblue] (3) to 
	node[draw=none, fill = none, font = \fontsizeEdge,
	below, yshift = 0.15cm, text = black] {$\Uc^4$}node[draw=none, fill = none, font = \fontsizeEdge,
	above, yshift = \yshiftLabel, text = black] {$c_{3,4}\left(\,\cdot\,,\,\cdot\,;\thetab_{3,4}\right)$} (4);    
	\draw[dashed, very thick, TUMblue] (1-2) to 
	node[draw=none, fill = none, font = \fontsizeEdge,
	below, yshift = 0.15cm, text = black] {$\Uc^3$, $\Uc^4$}node[draw=none, fill = none, font = \fontsizeEdge,
	above, yshift = \yshiftLabel, text = black] {$c_{1,3;2}\left(\,\cdot\,,\,\cdot\,;\thetab_{1,3;2}\right)$} (2-3);
	\draw[dotted, very thick, TUMlightblue] (2-3) to
	node[draw=none, fill = none, font = \fontsizeEdge,
	below, yshift = 0.15cm, text = black] {$\Uc^4$} node[draw=none, fill = none, font = \fontsizeEdge,
	above, yshift = \yshiftLabel, text = black] {$c_{2,4;3}\left(\,\cdot\,,\,\cdot\,;\thetab_{2,4;3}\right)$} (3-4);
	\draw[dotted, very thick, TUMlightblue] (1-3_2) to 
	node[draw=none, fill = none, font = \fontsizeEdge,
	below, yshift = 0.15cm, text = black] {$\Uc^4$} node[draw=none, fill = none, font = \fontsizeEdge,
	above, yshift = \yshiftLabel, text = black] {$c_{1,4;2,3}\left(\,\cdot\,,\,\cdot\,;\thetab_{1,4;2,3}\right)$} (2-4_3);
	
	\node[TreeLabels] (CT1)  [left of = 1, xshift = 0.50cm, font = \sf] {Tree 1}
	node[TreeLabels] (CT2)  [below of = CT1, yshift = \yshift, font = \sf] {Tree 2}
	node[TreeLabels] (CT3)  [below of = CT2, yshift = \yshift, font = \sf] {Tree 3}
	;			    
	\end{tikzpicture}
	\caption{Illustration of the four-dimensional D-vine describing the components of the dependence structure of the full model $c_{1:4}$ (dark, medium and light). The sub-vines for $c_{1:2}$ (dark) and $c_{1:3}$ (dark and medium) are highlighted by different color intensities of the nodes and line types of the edges. Above and below each edge the associated pair-copula and the observations that can be used for estimation are denoted, respectively.}
	\label{fig:dvine4d}
\end{figure}

If a continuous random vector $\Ub_{1:d}=(U_1,\ldots,U_d)^\top$ with uniform marginal distributions follows a D-vine copula density with order 1--2--\ldots--$d$, then, using the notation of \cite{czado2010}, the density can be written as
\begin{equation}\label{eq:dvinedensity}
\begin{split}
c_{1:d}(u_1,\ldots,u_d)=& \prod_{\ell=1}^{d-1} \prod_{k=1}^{d-\ell} c_{k,k+\ell;(k+1):(k+\ell-1)}\big(C_{k|(k+1):(k+\ell-1)}(u_{k}|u_{k+1},\ldots,u_{k+\ell-1}),\\
&C_{k+\ell|(k+1):(k+\ell-1)}(u_{k+\ell}|u_{k+1},\ldots,u_{k+\ell-1}); u_{k+1},\ldots,u_{k+\ell-1}\big).
\end{split}
\end{equation}
Here $c_{k,k+\ell;(k+1):(k+\ell-1)}(\,\cdot\,,\cdot\,;u_{k+1},\ldots,u_{k+\ell-1})$ is the bivariate copula density associated with the distribution of $(U_k,U_{k+\ell})^\top$ given $(U_{k+1},\ldots,U_{k+\ell-1})^\top=(u_{k+1},\ldots,u_{k+\ell-1})^\top$ and $C_{k|(k+1):(k+\ell-1)}(\,\cdot\,|u_{k+1},\ldots,u_{k+\ell-1})$ is the distribution function of the conditional distribution of $U_k$ given $(U_{k+1},\ldots,U_{k+\ell-1})^\top=(u_{k+1},\ldots,u_{k+\ell-1})^\top$, $\ell=1,\ldots,d-1$ and $k=1,\ldots,d-\ell$. The corresponding graphical interpretation is the tree representation, where the pair-copulas occurring in tree $j$ have a conditioning set of size $j-1$, $j=1,\ldots,d-1$. For $d=4$ this concept is illustrated in \autoref{fig:dvine4d}. 

Many authors make the so-called simplifying assumption that the pair-copulas $c_{k,k+\ell;(k+1):(k+\ell-1)}(\,\cdot\,,\cdot\,;u_{k+1},\ldots,u_{k+\ell-1})$ do not depend on values of the conditioning variables $u_{k+1},\ldots,u_{k+\ell-1}$. More detailed investigations of this assumption can for example be found in \cite{haff2010simplified}, \cite{acar2012beyond}, \cite{stoeber2013simplified}, \cite{spanhel2015simplified} and \cite{killiches2017examination}. We also make this assumption in order to ease inference later on although we could set up our model without it as well.

In the following we will assume a parametric model such that a D-vine copula can be identified by the set of pair-copula families $\Cc=(c_{k,k+\ell;(k+1):(k+\ell-1)}\mid k=1,\ldots,d-\ell\text{ and }\ell=1,\ldots,d-1)$ and the set of associated parameters $\thetab=(\thetab_{k,k+\ell;(k+1):(k+\ell-1)}\mid k=1,\ldots,d-\ell\text{ and }\ell=1,\ldots,d-1)$. In general, non-parametric pair-copulas could also be used \citep[see][]{nagler2016evading}.

A convenient property is that D-vine models are nested in the sense that the pair-copulas needed to describe the dependence of variables 1 to $j$ are contained in the model describing the dependence of variables 1 to $j+1$, $j<d$. This is illustrated in \autoref{fig:dvine4d}.
\paragraph{Model}\label{sec:model}\mbox{}\\
Since the data has been obtained from repeated measurements there exists a clear sequential or temporal ordering. This immediately suggests the use of D-vine copulas with order 1--2--\ldots--$d$ \citep{smith2010modeling, smith2015copula, NaiRuscone2017}. Therefore, as a general approach, we assume parametric simplified D-vine models (cf.\ \autoref{eq:dvinedensity}) for the copula densities of all groups $j=2,\ldots,d$. Of course, we only consider groups for which we have observations. The copula density $c_{1:j}^j$ of group $j$ then can be described with the help of the set of the $j(j-1)/2)$ pair-copula families 
\[
\Cc^j=(c_{k,k+\ell;(k+1):(k+\ell-1)}^j\mid k=1,\ldots,j-\ell\text{ and }\ell=1,\ldots,j-1)
\]
and the set of corresponding parameters 
\[
\thetab^j=(\thetab_{k,k+\ell;(k+1):(k+\ell-1)}^j\mid k=1,\ldots,j-\ell\text{ and }\ell=1,\ldots,j-1)
\]
for $j=2,\ldots,d$ with a non-empty $\Uc^j$. For the estimation of $\Cc^j$ and $\thetab^j$ we set up the likelihood, which is based on the subset of $\Uc$ containing the observations of length $j$. The resulting likelihood and log-likelihood can be written as
\[
L_j(\Cc^j,\thetab^j\mid\Uc^j)=\prod_{i\in I_j}c_{1:j}^j(u_1^i,\ldots,u_j^i\mid\Cc^j,\thetab^j)
\]
and 
\[
\log L_j(\Cc^j,\thetab^j\mid\Uc^j)=\sum_{i\in I_j}\log c_{1:j}^j(u_1^i,\ldots,u_j^i\mid\Cc^j,\thetab^j),
\]
respectively. Consequently, the log-likelihood of the general model is given by
\begin{equation}\label{eq:llgeneral}
\log L(\Cc^2,\ldots,\Cc^d, \thetab^2,\ldots,\thetab^d\mid\Uc)=\sum_{j=2}^d\log L_j(\Cc^j,\thetab^j\mid\Uc^j).
\end{equation}
For future reference we call this \emph{Model A}. It is obvious by construction that the models for different groups can be estimated independently from each other since there are no intersections between the groups, neither regarding data nor pair-copula families or parameters. From a practical point of view this would correspond to the assumption that the dependence structure of two groups can be completely different such that an individual for whom we have observed $j$ events have nothing in common with those who have had $j+1$ events. However, one can argue that an individual from group $j$ is basically a member of group $j+1$ for whom the $(j+1)$st measurement has not been observed yet. The underlying random mechanism (i.e.\ the copula), however, should be the same or at least share some properties. Therefore, it makes sense to impose more restrictions on the set of pair-copula families and the associated parameters. For example, one could assume that all groups share the same pair-copula families and only the parameters can differ between the groups. The most sensible and interesting case---which we will pursue in the following---is the one that all groups have the same pair-copula families and parameters, i.e.\ for all $j=2,\ldots,d$ we have 
\begin{equation}\label{eq:modelB}
\begin{split}
c_{k,k+j;(k+1):(k+j-1)}^j&=c_{k,k+j;(k+1):(k+j-1)},\\
\thetab_{k,k+j;(k+1):(k+j-1)}^j&=\thetab_{k,k+j;(k+1):(k+j-1)}
\end{split}
\end{equation} 
for $k=1,\ldots,j-\ell$ and $\ell=1,\ldots,j-1$. We will refer to this model as \emph{Model B}. Using the same families and parameters for all groups implies that the D-vine describing the dependence pattern of group $j$ is a sub-vine of the vine of groups $j+1,\ldots,d$. In particular, the vine copula density of group $j$ is simply the multivariate marginal density $c_{1:j}$ of the density $c_{1:d}$ of the largest group $d$. Consequently, $c_{1:d}$ describes the full model, from which the models of all smaller groups can be explicitly derived. Thus, the corresponding log-likelihood only depends on one set of $d(d-1)/2$ pair-copula families $\Cc=(c_{k,k+\ell;(k+1):(k+\ell-1)}\mid k=1,\ldots,d-\ell\text{ and }\ell=1,\ldots,d-1)$ and the set of corresponding parameters $\thetab=(\thetab_{k,k+\ell;(k+1):(k+\ell-1)}\mid k=1,\ldots,d-\ell\text{ and }\ell=1,\ldots,d-1)$. 

\paragraph{Example}\mbox{}\\
In order to illustrate the above concept we will now look at the example with at most $d=4$ repeated measurements. Assume we have (up to) four-dimensional repeated measurement data $\Uc=\left\{\ub^1,\ldots,\ub^n\right\}$ of size $n=n_2+n_3+n_4$ ordered as described in \autoref{sec:setting}, which can be partitioned into groups $2$, $3$ and $4$ by defining $\Uc^j=\left\{u^i\mid i\in I_j\right\}$, $j=2,3,4$, where $I_2=\left\{i\mid \ub^i\in\Rbb^2\right\}$, $I_3=\left\{i\mid \ub^i\in\Rbb^3\right\}$ and $I_4=\left\{i\mid \ub^i\in\Rbb^4\right\}$. The model and hence the log-likelihood depends on the set of the six pair-copulas $\Cc=(c_{1,2}, c_{2,3}, c_{3,4}, c_{1,3;2}, c_{2,4;3}, c_{1,4;2,3})$ and the associated parameters $\thetab=(\thetab_{1,2}, \thetab_{2,3}, \thetab_{3,4}, \thetab_{1,3;2}, \thetab_{2,4;3}, \thetab_{1,4;2,3})$. \autoref{fig:dvine4d} shows a schematic representation of the full model $c_{1:4}$ with its pair-copulas and parameters. The nodes represent the measurements. Above and below each edge the associated pair-copula and the observations that can be used for estimation are denoted, respectively. The sub-vines for $c_{1:2}$ and $c_{1:3}$ are highlighted by different color intensities of the nodes and line types of the edges. The resulting log-likelihood is given by
\begin{equation}\label{eq:4ll_1}
\begin{split}
\log L(\Cc,\thetab\mid \Uc)=&\sum_{i\in I_2}\log c_{1:2}(u_1^i,u_2^i\mid c_{1,2},\thetab_{1,2})\\
&+\sum_{i\in I_3}\log c_{1:3}(u_1^i,u_2^i,u_3^i\mid c_{1,2},c_{2,3}, c_{1,3;2},\thetab_{1,2}, \thetab_{2,3}, \thetab_{1,3;2})\\
&+\sum_{i\in I_4}\log c_{1:4}(u_1^i,u_2^i,u_3^i,u_4^i\mid c_{1,2},c_{2,3}, c_{3,4}, c_{1,3;2},c_{2,4;,3}, c_{1,4;2,3},\\
&\phantom{+\sum_{i\in I_4}\log c_{1:4}(u_1^i,u_2^i,u_3^i,u_4^i\mid \mbox{}}\thetab_{1,2}, \thetab_{2,3}, \thetab_{3,4}, \thetab_{1,3;2}, \thetab_{2,4;,3}, \thetab_{1,4;2,3})
\end{split}
\end{equation}
Using the vine decomposition from \autoref{eq:dvinedensity} for $c_{1:2}$, $c_{1:3}$ and $c_{1:4}$, the log-likelihood associated with data $\Uc$ (\autoref{eq:4ll_1}) can be re-written as
\begin{equation}\label{eq:4ll_2}
\begin{split}
\log &\, L(\Cc,\thetab\mid \Uc)=\\
&\sum_{i\in I_2\cup I_3 \cup I_4}\log c_{1,2}(u_1^i,u_2^i; \thetab_{1,2})\\
&+\sum_{i\in I_3\cup I_4}\Big[ \log c_{2,3}(u_2^i,u_3^i; \thetab_{2,3})+ \log c_{1,3;2}(C_{1|2}(u_1^i|u_2^i;\thetab_{12}),C_{3|2}(u_3^i|u_2^i;\thetab_{23}); \thetab_{1,3;2})\Big]\\
&+\sum_{i\in I_4}\Big[ \log c_{3,4}(u_3^i,u_4^i;\thetab_{3,4})+ \log c_{2,4;3}(C_{2|3}(u_2^i|u_3^i;\thetab_{2,3}),C_{4|3}(u_4^i|u_3^i;\thetab_{4,3}); \thetab_{2,4;3})\Big.\\
& \phantom{+\sum_{i\in I_4}\big[ \big.} + \log c_{1,4;2,3}(C_{1|3;2}(C_{1|2}(u_1^i|u_2^i;\thetab_{1,2})\,|\,C_{3|2}(u_3^i|u_2^i;\thetab_{2,3}); \thetab_{1,3;2}),\\
&\Big. \phantom{+\sum_{i\in I_4}\big[ \big.+\log c_{1,4;2,3}\left( \right.}C_{4|2;3}(C_{2|3}(u_2^i|u_3^i;\thetab_{2,3})\, | \, C_{4|3}(u_4^i|u_3^i;\thetab_{4,3}); \thetab_{2,4;3}); \thetab_{1,4;2,3})\Big]
\end{split}
\end{equation}

For the general case of Model A (\autoref{eq:llgeneral}) we saw that the pair-copulas and parameters corresponding to group $j$ can be estimated independently from those of the remaining groups and only depend on the data contained in $\Uc^j$. Looking at \autoref{eq:4ll_2} (corresponding to Model B) it immediately becomes clear that assuming the pair-copulas and parameters are the same for all groups has changed this phenomenon. The D-vines describing the densities $c_{1:2}$ and $c_{1:3}$ are nested sub-vines of the full model $c_{1:4}$, which can easily be understood from \autoref{fig:dvine4d}: The dark nodes (and solid edges) correspond to $c_{1:2}$; adding the medium colored nodes (and dashed edges) results in the model of $c_{1:3}$; incorporating also the light nodes (and dotted edges) yields the full model for $c_{1:4}$. Therefore, when it comes to estimation we see for example that not only the observations belonging to $\Uc^2$ but also those from $\Uc^3$ and $\Uc^4$ (i.e.\ the entire sample $\Uc$) have an influence on the estimate $c_{1,2}$ and $\thetab_{1,2}$. Thus this increases the accuracy of the estimation compared to the approach from Model A. 

The assumption of common pair-copula families and parameters for all groups come with the advantages of better interpretability, less parameters and higher estimation accuracy. 

\paragraph{Missing values}\label{sec:missingvalues} \mbox{}\\ 
In practice, unfortunately, data do not always look exactly the way we described it in \autoref{sec:setting}. Sometimes there are missing values in the data. For example, there might be individuals for whom the first, third, fourth and fifth measurement are available but the second one is missing. Such situation can occur for various reasons, e.g.\ a patient skips a measurement date due to illness, measuring instruments have problems causing a loss of the result or data is simply not reported due to human failure. Moreover, there might be (non-informative) dropouts, i.e.\ individuals with measurements only up to a certain time, e.g.\ caused by relocation of a patient to another city. For many model classes such observations cannot be used at all and have to be removed from the data set for model estimation. This way the sample size is decreased and information is lost. For Model B, however, observations with missing values can still be used (assuming they are missing at random). The information gained from our exemplary individual with measurements 1, 3, 4, 5 still contributes to the estimation of $c_{3,4}$, $c_{4,5}$ and $c_{3,5;4}$ (and of course to the estimation of the marginal distributions 1, 3, 4, 5). Since the missing second measurement is needed for the estimation of the remaining pair-copulas, this individual cannot be used in order to estimate them. Nevertheless, we prevent the loss of the individual's entire information. In order to include observations with missing value into our model we simply have to modify the log-likelihood such that the sums of the log-likelihood of each pair-copula includes all observations for whom the necessary measurements are available. For the sake of notation we will stick to the formulation of Model B as above, keeping in mind that missing values can also be handled by our approach.

\paragraph{Conditional prediction}\label{sec:quantileprediction} \mbox{}\\ 
Further, we can use our repeated measurement data model for prediction. In many applications it can be interesting to have a prediction for the size of an upcoming measurement. For instance, having proper estimates for future claims can be a competitive advantage for the risk management department of an insurance company. 

For a $d$-dimensional model, consider an individual $i$ that has had $d_i<d$ measurements so far, i.e.\ $\yb^i=(y_1^i,\ldots,y_{d_i}^i)^\top$. We are now interested in the distribution of the next measurement $d_i+1$. Since $d_i+1\leq d$, the sub-vine describing the dependence of events 1 to $d_i+1$ can be extracted from the full model. We consider the conditional distribution function $F^i_{d_i+1|1:d_i}(\,\cdot\,|y^i_1,\ldots,y^i_{d_i})$. \cite{joe1997multivariate} was the first to show that there exists a recursive representation for such conditional distribution functions. This way one obtains a closed-form expression of the conditional distribution function solely based on the pair-copulas specified in the D-vine (and the univariate marginals, of course) if the variable to be predicted is a leaf in the first tree. In our case, $d_i+1$ is in fact a leaf in the first tree of the D-vine on nodes 1 to $d_i+1$. Thus, we know $F^i_{d_i+1|1:d_i}(\,\cdot\,|y^i_1,\ldots,y^i_{d_i})$ analytically and can further simulate from it. For example, we can express $F^i_{4|1,2,3}$ in the following way:
\[
\begin{split}
F^i_{4|1,2,3}(y_4^i|y_1^i,y_2^i,y_3^i)=C_{4|1;23}&\left(C_{4|2;3}\left(C_{4|3}(F^i_4(y^i_4)|F^i_3(y^i_3))\big|C_{2|3}(F^i_2(y^i_2)|F^i_3(y^i_3))\right)\big|\right.\\
&\phantom{\big(\big.}\left. C_{1|3;2}\left(C_{1|2}(F^i_1(y^i_1)|F^i_2(y^i_2))\big|C_{3|2}(F^i_3(y^i_3)|F^i_2(y^i_2))\right)\right).
\end{split}
\]
Further, the conditional quantile function can be expressed in general as
\begin{equation}\label{eq:condquantile}
\begin{split}
q_{\alpha}(y^i_1,\ldots,y^i_{d_i}) &= (F^i_{d_i+1|1:d_i})^{-1}(\alpha|y^i_1,\ldots,y^i_{d_i}) \\
&= (F^i_{d_i+1})^{-1}\left(C_{d_i+1|1:d_i}^{-1}(\alpha|F^i_1(y^i_1),\ldots,F^i_{d_i}(y^i_{d_i}))\right)
\end{split}
\end{equation}
and is of great interest in order to determine upper and lower bounds of a confidence interval. \cite{kraus2017d} show that inversion also yields a closed-form expression for the conditional quantile function solely based on the specified pair-copulas and marginals. Thus, we can determine arbitrary conditional quantiles for the size of measurement $d_i+1$. For example, for financial applications it might be interesting to obtain a conditional $99\%$-value at risk, i.e.\ the conditional $99\%$-quantile, for the size of individual $i$'s next measurement.\\

In order to be able to perform statistical inference of any kind with our D-vine model we first have to estimate the pair-copula families and associated parameters. \autoref{sec:estimation} will present two estimation approaches. First, however, we will introduce linear mixed model and illustrate how they are connected to our proposed D-vine based model in \autoref{sec:connectionLMM}.

\section{Connection between the D-vine based model and linear mixed models}\label{sec:connectionLMM}
Probably the most popular models for longitudinal data are linear mixed models. In this section we will give a short introduction to this model class and show how they are connected to our approach from \autoref{sec:modeling}.

\subsection{Linear mixed models for repeated measurements}\label{sec:LMM} 
Linear mixed models have been discussed in detail by many authors, e.g.\ in \cite{diggle2002analysis}, \cite{verbeke2009linear} and \cite{fahrmeir2013regression}. Describing the outcome of repeated measurements $j$, $j=1,\ldots,d_i$, for individuals $i$, $i=1,\ldots,n$ as responses $Y^i_j$, they extend linear models by including random effects $\gammab_i\in\Rbb^q$ to the fixed (i.e.\ non-random) effects $\beta\in\Rbb^p$, $p,q\in \Nbb$. These random effects, unlike the fixed effects, are different for each individual. The covariate vectors $\xb_{i,j}\in\Rbb^p$ and $\zb_{i,j}\in\Rbb^q$ are associated to the fixed and random effects, respectively.

For $i=1,\ldots,n$ and $j=1,\ldots,d_i$, the $j$th measurement for individual $i$ is assumed to decompose to
\begin{equation}\label{eq:LMMobs}
Y^i_{j}=\xb_{i,j}^\top\betab+ \zb_{i,j}^\top\gammab_i +\eps_{i,j},
\end{equation}
where the vector of random effects $\gammab_i\sim\Nc_q(\0,D)$ is normally distributed with zero expectation covariance matrix $D\in\Rbb^{q\times q}$ and the error vector $\epsb_i=(\eps_{i,1}, \ldots, \eps_{i,d_i})^\top\sim\Nc_{d_i}(\0,\Sigma_i)$ also follows a centered normal distribution with covariance matrix $\Sigma_i\in \Rbb^{d_i\times d_i}$. Further, $\gammab_1,\ldots,\gammab_n$, $\epsb_1,\ldots,\epsb_n$ are assumed to be independent. Hence,
\begin{equation}\label{eq:LMMdistribution}
Y^i_j\sim \Nc(\xb_{i,j}^\top\betab,\phi_{i,j}^2)
\end{equation}
with standard deviation $\phi_{i,j}:=\left(\zb^\top_{i,j}D\zb_{i,j}+\sigma^2_{i,j}\right)^{1/2}$, where $\sigma^2_{i,j}:=\Var(\eps_{i,j})$. Using the notation
\[
X_i:=\begin{pmatrix}
\xb_{i,1}^\top\\
\vdots\\
\xb_{i,d_i}^\top
\end{pmatrix}\in\Rbb^{d_i\times p},\quad  
Z_i:=\begin{pmatrix}
\zb_{i,1}^\top\\
\vdots\\
\zb_{i,d_i}^\top
\end{pmatrix}\in\Rbb^{d_i\times q}, \quad 
\Yb^i:=\begin{pmatrix}
Y^i_1\\
\vdots\\
Y^i_{d_i}
\end{pmatrix}\in\Rbb^{d_i}
\]
we can represent the vector of all measurements belonging to individual $i$ as follows:
\begin{equation}\label{eq:LMMvec}
\Yb^i=X_{i}\betab+ Z_{i}\gammab_i +\epsb_i.
\end{equation}
We see that due to the independence assumptions of $\gammab_i$ and $\epsb_i$, $i=1,\ldots,n$, there exists a correlation between measurements of one individual but measurements of different individuals are independent. Further, the joint distribution of $\Yb^i$ can be determined to be
\begin{equation}\label{eq:LMMvecdistribution}
\Yb^i\sim \Nc_{d_i}(X_i\betab, Z_iDZ_i^\top+\Sigma_i)
\end{equation}
and $\Yb^1,\ldots, \Yb^n$ are independent. The fixed effects $\betab$ and random effects $\gammab_i$ as well as the parameters of the covariance matrices $D$ and $\Sigma_i$, $i=1,\ldots,n$, can be estimated using maximum likelihood estimation as described for example in \cite{diggle2002analysis} and \cite{fahrmeir2013regression}.

Linear mixed models are very popular in practice since they are easy to handle and interpret. Further, observations with missing data can also be used for ML estimation as long as the values are missing at random \citep[see e.g.][]{mcculloch2011generalized,ibrahim2009missing}.

\subsection{Aligning linear mixed models and the D-vine based approach}\label{sec:} 
\autoref{eq:LMMvecdistribution} implies that all univariate marginal distributions are normal distributions. Further, the dependence structure is Gaussian and can vary from individual to individual since the correlation matrix $R_i$ of $\Yb^i$ is given by
\[
R_i:=\Corr(\Yb^i)=\diag(\phi_{i,1}^{-1},\ldots,\phi_{i,d_i}^{-1})\left(Z_iDZ_i^\top+\Sigma_i\right)  \diag(\phi_{i,1}^{-1},\ldots,\phi_{i,d_i}^{-1}), 
\]
where $\phi_{i,j}$ is the standard deviation of $Y^i_j$, $j=1,\ldots,d_i$, $i=1,\ldots,n$. In practice, however, this would make estimation infeasible since the number of parameters would be too large; in many cases one would even have more parameters than observations. Therefore, structural assumptions are made, especially for $\Sigma_i\in \Rbb^{d_i\times d_i}$, in order reduce the number of parameters to be estimated. 

In \autoref{sec:dependencemodel} we assumed that the dependence structure is basically the same for all individuals and only differs due to the number of measurements $d_i$ that individual $i$ has had so far. In order to obtain the same for linear mixed models, we simply have to require the following homogeneity condition:\\[3mm]
\textsf{\textbf{Homogeneity condition:}} We call correlation matrices $R_i$ \emph{homogeneous} if they are the same for all individuals $i=1,\ldots,n$ except for the dimension, i.e.\  $R_i=(r_{k,\ell})_{k,\ell=1}^{d_i}\in\Rbb^{d_i\times d_i}$ is a $(d_i\times d_i)$-submatrix of a correlation matrix $R=R_d=(r_{k,\ell})_{k,\ell=1}^{d}\in\Rbb^{d\times d}$. \\[3mm]
This condition is in particular fulfilled if the covariance matrices of the errors $\Sigma_i\in\Rbb^{d_i\times d_i}$ and the design matrices of the random effects $Z_i\in\Rbb^{d_i\times q}$ are constant in $i$ except for the dimension. Despite being a restriction, linear mixed models meeting this requirement still comprise a wide range of models used in practice. The assumption on the covariance matrices $\Sigma_i$ is for example fulfilled if errors
\begin{itemize}\label{list:corstructures}
	\item are assumed to be i.i.d., i.e.\ the $(k,\ell)$th entry of $\Sigma_i$ is given by $\sigma^2\1\{k=\ell\}$, where $\1\{\cdot \}$ denotes the indicator function;
	\item exhibit a compound symmetry structure, i.e.\ the $(k,\ell)$th entry of $\Sigma_i$ is  $\sigma^2\rho^{\1\{k\neq \ell\}}$ for some $\rho\in(-1,1)$;
	\item follow an autoregressive structure of order 1 (AR(1)), i.e.\ the $(k,\ell)$th entry of $\Sigma_i$ is given by $\sigma^2\rho^{\left|k-\ell\right|}$ for some $\rho\in(-1,1)$;
	\item have an exponential decay structure, i.e.\ the $(k,\ell)$th entry of $\Sigma_i$ is given by $\sigma^2\exp\left\{-\left|k-\ell\right|/r\right\}$, where $r>0$ is the constant ``range'' parameter.
\end{itemize}
These are typical simplifications that are made anyway for modeling longitudinal data in most applications if the number of individuals is large with respect to the number of measurements. The assumption on the design matrices $Z_i$ is also often satisfied, e.g.\ for the popular class of so-called \emph{random intercept models}, where $Z_i=(1,\ldots,1)^\top\in\Rbb^{d_i\times 1}$ for $j=1,\ldots, d_i$ and $i=1,\ldots,n$. Further, the assumption includes any model where the covariates associated with the random effect only depend on the (common) measurement times $t_j$, $j=1,\ldots,d$, i.e.\ for example $Z_i=(t_1,\ldots,t_{d_i})^\top\in\Rbb^{d_i\times 1}$ or more generally $Z_i=(h(t_1),\ldots,h(t_{d_i}))^\top\in\Rbb^{d_i\times 1}$ for some function $h\colon \Rbb\to\Rbb$. Thus, assuming that $Z_i$ only depends on the number of measurements $d_i$ for individual $i$ is also not uncommon such that there is in fact a wide class of linear mixed models sharing the property that the correlation matrix $R_i$ of $\Yb^i$ only depends on the number of measurements.

If $R_i$ is homogeneous in $i$, we have that all individuals $i$ share the same Gaussian dependence structure, i.e.\ correlation matrix. This scenario is a special case of the D-vine based model since we can represent any Gaussian correlation matrix using a D-vine with Gaussian pair-copulas and the corresponding (partial) correlations as parameters \citep[see for example][Theorem 4.1]{stoeber2013simplified}. The univariate margins $F^i_j$ can be chosen arbitrarily for the copula approach such that we can simply use $\Nc(\xb_{i,j}^\top\betab,\phi_{i,j}^2)$-margins (cf.\ \autoref{eq:LMMdistribution}) to end up with a model describing the same joint distribution of $\Yb^i$ as the corresponding linear mixed model (\autoref{eq:LMMvecdistribution}). Since we can use arbitrary distributions for the margins and/or any D-vine copula for the dependence structure, our approach can be seen as an extension of linear mixed models with common correlation structure for all individuals. \autoref{fig:flowchartLMM_Dvine} illustrates the link between our D-vine based model and linear mixed models.

\tikzstyle{block} = [rectangle, draw=TUMdarkblue, fill=TUMlightblue, 
text width=10em, text centered, rounded corners, minimum height=4em,
text = white, font = \bfseries\sffamily, very thick]
\tikzstyle{line} = [draw, -latex']
\tikzstyle{DummyNode}  = [draw = none, fill = none, text = white]
\begin{figure}[H]
	\centering
	\begin{tikzpicture}[node distance = 3.2cm]
	\node [block] (LMM) {Linear mixed model};
	\node [DummyNode, left of=LMM] (dummy11) {};
	\node [DummyNode, below of=dummy11] (dummy21) {};
	\node [DummyNode, right of=LMM] (dummy13) {};
	\node [DummyNode, below of=dummy11] (dummy23) {};
	\node [rectangle split, rectangle split parts=2, draw=TUMdarkblue, fill=TUMlightblue, 
	text width=10em, text centered, rounded corners, minimum height=4em,
	text = white, font = \bfseries\sffamily, very thick, below of=LMM] (LMMconstcorr) {LMM with common correlation structure for all individuals
		\nodepart{two} Gaussian copula with Gaussian regression margins};
	\node [DummyNode, below of=LMMconstcorr] (dummy33) {};
	\node [block, yshift = -1em, left of=dummy33] (GaussianArb) {Gaussian copula with arbitrary margins};
	\node [block, yshift = -1em, right of=dummy33] (DvineGauss) {D-vine copula with Gaussian regression margins};
	\node [block, below of=dummy33] (DvineArb) {D-vine copula with \\arbitrary margins};
	\path [very thick, TUMdarkblue] (LMM) edge (LMMconstcorr);
	\path [very thick, TUMdarkblue] (LMMconstcorr) edge (GaussianArb)
	edge (DvineGauss);
	\path [very thick, TUMdarkblue] (DvineGauss) edge (DvineArb);
	\path [very thick, TUMdarkblue] (GaussianArb) edge (DvineArb);
	\end{tikzpicture}
	
	\caption{Flow chart illustrating how the D-vine based model is linked to linear mixed models.}
	\label{fig:flowchartLMM_Dvine}
\end{figure}
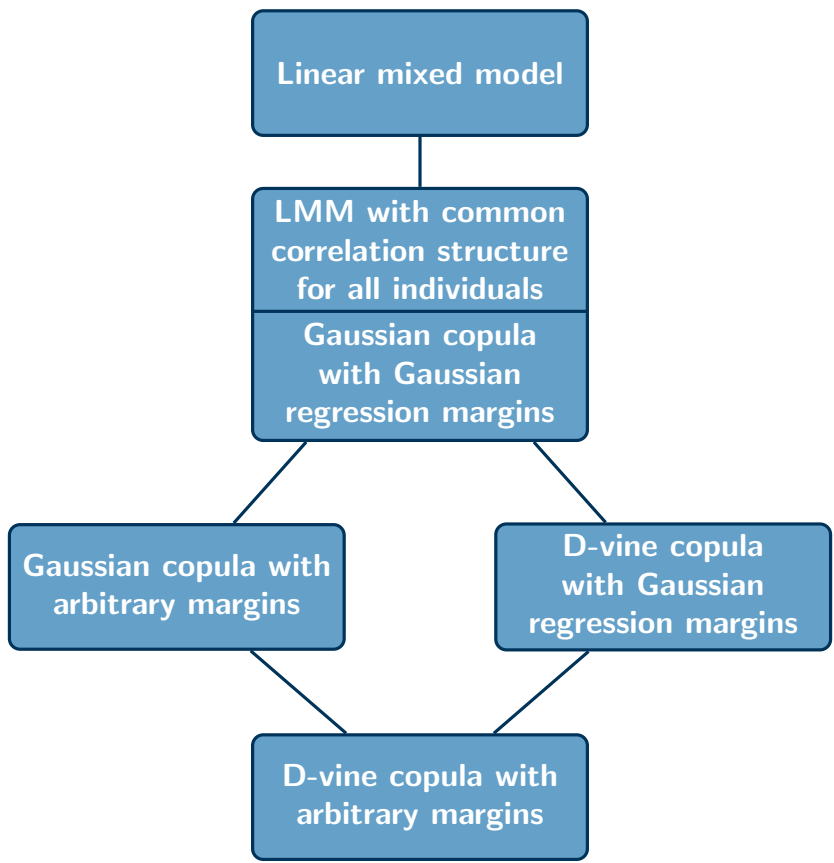
For the application in \autoref{sec:application} we will compare how well both model classes perform fitting real life data.

\section{Estimation methods for the D-vine based model}\label{sec:estimation}

\subsection{Marginal modeling}\label{sec:marginalmodeling}
Although our focus is on dependence modeling, we will briefly discuss how the univariate marginal models for $Y^i_j$, i.e.\ $F^i_j$, can be estimated. In general the choice of marginal models is completely arbitrary. They can be parametric or non-parametric. The most common situation for repeated measurements is that in addition to the measurement data $\Yc$ itself further covariates are known for each individual and measurement. Therefore, regression models such as linear (LMs) or generalized linear models (GLMs) can be applied. In this case, $F^i_j(\, \cdot \, )=F_j(\,\cdot\, | \, \xb_{i,j})$ depends on the individual's associated covariates $\xb_{i,j}\in\Rbb^p$, where $p\in \Nbb$ is the number of covariates used in the model. In our application (\autoref{sec:application}) we will fit linear models to the margins. In order to avoid overfitting we determine which covariates to include into the models by BIC-based forward selection, i.e.\ covariates are added to the model until BIC \citep[Bayesian information criterion, see][]{schwarz1978estimating} cannot be improved anymore. Our focus, however, is rather to develop a flexible model describing the dependence structure that is present in the data $\Yc$ such that we will not further elaborate on how to estimate the univariate marginal distributions.

\subsection{Dependence modeling}\label{sec:depmodeling}
Assume we have estimated the marginal distributions and obtained (pseudo-)copula data by applying the estimated distribution functions $\hat{F}^i_j$ to the measurements, i.e.\ $\hat{u}_j^i:=\hat{F}^i_j(y_j^i)$. We now use the transformed data as a copula sample to estimate the underlying dependence structure. \autoref{sec:modeling} has shown that D-vine copulas are suited for modeling the dependence structure being present in repeated measurement data. Model B (\autoref{eq:modelB}) was preferable since it is easier to interpret and estimate. Further predictions for not yet observed measurements can be made. The aim of the methods presented in \autoref{sec:estimation} is to find estimates for the set of pair-copula families $\Cc=(c_{k,k+\ell;(k+1):(k+\ell-1)}\mid k=1,\ldots,d-\ell\text{ and }\ell=1,\ldots,d-1)$ and the set of parameters $\thetab=(\thetab_{k,k+\ell;(k+1):(k+\ell-1)}\mid k=1,\ldots,d-\ell\text{ and }\ell=1,\ldots,d-1)$ corresponding to Model B from \autoref{sec:modeling}, where $d$ is the maximal number of observed events per observation. We will present two approaches: a standard joint maximum-likelihood estimator and a sequential method. Since we want to choose both parameters and families for each pair-copula we will select from a set of $m$ bivariate candidate family types $\Gamma=\left\{\gamma_1,\ldots,\gamma_m\right\}$, where each member $\gamma\in\Gamma$ has its own space of admissible parameters $\Omega(\gamma)$.

\paragraph{Joint maximum-likelihood approach}\mbox{}\\
The canonical approach in order to find optimal pair-copula families and parameters would be to use maximum-likelihood estimation. In \autoref{sec:modeling} we have already determined the log-likelihood. Since the families specify which parameters are admissible, finding the optimal families and parameters is divided into two steps: For each combination of families we have to determine the maximum-likelihood estimate of parameters; then we select the one combination with the overall highest likelihood. This way we find the best D-vine model with regards to likelihood optimization. However, since there are $\left|\Gamma\right|=m$ candidates for each of the $d(d-1)/2$ pair-copula families, we have to perform $m^{d(d-1)/2}$ times an at least $(d(d-1)/2)$-dimensional optimization (some families like the t-copula may have more than one parameter). It is obvious that this can very quickly become computationally infeasible if the number of candidate families $m$ is high and---especially---if the dimension $d$ gets large. \\

Of course, the possibly large number of parameters to be estimated is a general problem in the statistical analysis of vine copulas. Therefore, \cite{aasczado} (for D-vines) and later \cite{dissmann2013selecting} (for general vine copulas) developed a sequential tree-by-tree selection algorithm facilitating vine copula model estimation up to very high dimensions. Di\ss mann's algorithm is commonly used to fit the vine's model structure, pair-copula families and parameters but it can also only be used for the selection of families and parameters only if we have a fixed tree structure (e.g.\ a D-vine). The difference to the classical situation which we face when we want to estimate a vine copula is that our observations have different lengths. 

\paragraph{Sequential approach}\mbox{}\\
Inspired by Di\ss mann's algorithm we want to fit the pair-copula families and the associated parameters of the D-vine to a repeated measurement data set using a sequential approach. Given classical data, Di\ss mann's algorithm starts with the estimation of the first tree and estimates the unconditional pair-copulas (and their parameters) via maximum-likelihood estimation. Then the observations are transformed into pseudo-observations needed for the estimation of the second tree using the estimated pair-copulas of tree 1. Continuing this way the vine is built up tree-by-tree. 

In the presence of repeated measurement data, however, we can pursue a very similar strategy. The only difference is that we estimate each pair-copula (and its parameter(s)) only based on the available full observation. All pair-copulas to be estimated are of the form $c_{k,\ell;(k+1):(\ell-1)}$ with parameter $\thetab_{k,\ell;(k+1):(\ell-1)}$. When all observations are of the form $(u_1,\ldots,u_j)^\top$, i.e.\ there are no ``gaps'' between two observed events, we can use the information of observations with a minimum length of $\ell$, i.e.\ all observations in $\bigcup_{j=\ell}^d\Uc^j$, for the estimation of $c_{k,\ell;(k+1):(\ell-1)}$ and $\thetab_{k,\ell;(k+1):(\ell-1)}$. Thus, we can maintain the basic scheme known from Di\ss mann's algorithm. With a slight modification of the data we are even able to use the function \texttt{RVineCopSelect} from the R library \texttt{VineCopula} \citep{VC} for our purpose, making our approach also very appealing from a practitioner's point of view. 

Of course, this sequential approach can also be applied for data with missing values (\autoref{sec:dependencemodel}, \autopageref{sec:missingvalues}). Then, for the estimation of each pair-copula $c_{k,\ell;(k+1):(\ell-1)}$ with associated parameter $\thetab_{k,\ell;(k+1):(\ell-1)}$ is performed using all observations for whom the necessary measurements $u_k,u_{k+1},\ldots,u_l$ are available. The function \texttt{RVineCopSelect} can still be used in the presence of missing values. 

The biggest advantage of being able to use sequential estimation approach is that we can estimate models at reasonable computational costs, even in high dimensions. Of course, the approach also works when using non-parametric pair-copulas or even non-simplified vine copulas. For details for estimating non-parametric and non-simplified vines we refer the reader to \cite{nagler2016evading} and \cite{vatter2016gamvine}, respectively. Yet, as already mentioned at the beginning, we focus on parametric simplified vine copulas here.

\subsection{Model selection}
In model selection one often wants to compare different fitted models. For this purpose the log-likelihood and log-likelihood based measures such as AIC \citep{akaike1998information} and BIC \citep{schwarz1978estimating}, which penalize large numbers of parameters, are frequently applied. Whereas the penalty of the AIC only depends on the number of parameters in the model, that of BIC also depends on the sample size. In our case, however, it is not completely obvious what sample size to use. Therefore, we derive how the BIC for the D-vine based model including margins can be calculated in our situation. \autoref{prop:BIC} shows that each parameter is to be weighted with the (logarithm of) the number of observations that directly contribute to its estimation. A proof can be found in \appref{App:proof_prop}.
\begin{prop}\label{prop:BIC}
	Let $p_j\in\Nbb$ be the number of parameters of the D-vine based model including margins restricted to the measurements 1 to $j$, $j=1,\ldots,d$, and define $\Delta p_j:=p_j-p_{j-1}$ for $j=2,\ldots,d$ and $\Delta p_1:=p_1$. Further denote by $N_j=\sum_{k=j}^d n_k$ the number of individuals with at least $j$ measurements. The BIC of the D-vine based model including margins is given by
	\[
	\BIC=-2\log L(\hat{\thetab}\,|\,\Yc)+\sum_{j=1}^d\Delta p_j\log(N_j).
	\]	
	Here, $\log L(\hat{\thetab}\,|\,\Yc)=\log L(\hat{\thetab}_M\,|\,\Yc)+\log L(\hat{\thetab}_C\,|\,\Uc)$ is the log-likelihood of the fitted model including margins, i.e.\ the sum of the log-likelihood of the margins $\log L(\hat{\thetab}_M\,|\,\Yc)$ and the one of the copula $\log L(\hat{\thetab}_C\,|\,\Uc)$ (which is the one of Model B from \autoref{sec:dependencemodel}). Further, $\hat{\thetab}=(\hat{\thetab}_{M},\hat{\thetab}_{C})$ is the maximum-likelihood estimate for the set of all model parameters (associated with both the margins $\hat{\thetab}_{M}$ and the D-vine copula $\hat{\thetab}_{C}$).
\end{prop}

\begin{rem}
Although this BIC adjustment was developed for the D-vine based model, it can also be used for certain types of linear mixed models due to the connection described in \autoref{sec:connectionLMM}. For LMMs fulfilling the homogeneity condition the BIC can be determined with the formula from \autoref{prop:BIC} if only individuals with $j$ or more measurements contribute to the estimation of the parameters of the sub-model restricted to the first $j$ measurements which were not already contained in the sub-model restricted to the first $j-1$ measurements. This is for example the case if on the one hand no structural assumptions (besides homogeneity) are imposed on the covariance matrices of the random effects and the errors $D$ and $\Sigma_i$ and on the other hand the design matrices $X_i$ have a form that allows for different marginal regression models for different measurements. For guaranteeing the latter each covariate is only allowed to be incorporated in one of the marginal regressions, i.e.\ the values of this covariate are zero for all other measurements; if a covariate still is to be included in more than one model, one simply splits up the covariate into several measurements-specific covariates that are non-zero only for one particular measurement. This way an own coefficient for one covariate can be estimated for different marginal models (if necessary).
\end{rem}
\section{Simulation study}\label{sec:simstudy}
In order to check that the sequential estimation approach from \autoref{sec:estimation} works reasonably well, we perform a simulation study that is inspired by the data analyzed in \autoref{sec:application}. 
\paragraph{Simulation setting}\mbox{}\\
For a maximum number of measurements $d\in\{5,10\}$, we generate $d$-dimensional data sets and prune them randomly to obtain an unbalanced setting. In this context pruning means that for each $d$-dimensional observation $i$ we independently draw $d_i$ from a discrete distribution on $\{2,\ldots,d\}$ and restrict this observation to its first $d_i$ components. This way we mimic the nature of unbalanced repeated measurement data. In order to assess the implications of having only incomplete data we sequentially fit a D-vine copula to both the full and the pruned data set and compare the estimates. 

To obtain data sets we consider randomly generate D-vine copulas with structure 1--2--\ldots--$d$. For this purpose, we rely on the method proposed in \cite{joe2006generating} to sample Gaussian correlation matrices that are uniformly distributed over the space of valid correlation matrices. Conveniently, this method is already based on a vine decomposition: For each tree $i$, $i=1,\ldots,d-1$, we generate the corresponding $d-i$ parameters associated to the Gaussian pair-copulas by drawing from a $\text{Beta}((d-i+1)/2, (d-i+1)/2)$ distribution and transforming the outcome linearly to $[-1, 1]$, resulting in a mean and mode of $0$ and a variance of $1/(d-i+2)$. However, since we do not only want to consider Gaussian D-vines, we transform the Gaussian parameters to Kendall's $\tau$ values using the relationship $\tau=\frac{2}{\pi}\arcsin(\rho)$. Then, we randomly draw a pair-copula family for each pair-copula to be specified\footnote{The families are drawn uniformly from the ones available in the library \texttt{VineCopula}: Gaussian, t, Clayton, Gumbel, Frank, Joe, BB1, BB6, BB7, BB8 and Tawn as well their rotations \citep[see][for details]{VC}.} and convert the Kendall's $\tau$ values to parameters of the respective families. For one-parametric families $\tau$ can directly be transformed to the parameter space. For two-parametric families there are infinitely many combinations of parameters resulting in the same Kendall's $\tau$ value. Therefore, we adopt the approach used in \cite{kraus2017growing}: draw the second parameter randomly\footnote{The specific sampling distributions can be found in Appendix B of \cite{kraus2017growing}.} and determine the first parameter implicitly such that the two parameters imply the required Kendall's $\tau$. 

With the above procedure we generate $R=1000$ D-vine copulas and simulate data sets of size $n\in\{200,2000\}$. Then for each observation $i$ we randomly determine its length $d_i\in\{2,\ldots,d\}$. For $d=5$, the underlying distribution mimics the observed measurement rates of the data considered in \autoref{sec:application}. The exact proportions of individuals with a least $j$ measurements would have been $100.0\%, 78.5\%, 58.5\%, 43.9\%$ for $j=2,3,4,5$, respectively. For $d=10$, we extended the scenario of $d=5$ accordingly. The distributions are given in \autoref{tab:sim_study_freq}. 
\begin{table}[h!]
	\centering
	\begin{tabular}{ccccc}
		$j$ & 2 & 3 & 4 & 5 \\
		\hline
		probability of $d_i=j$ & 20\% & 20\% & 15\% & 45\% \\
		probability of $d_i\geq j$ & 100\% & 80\% & 60\% & 45\% \\
	\end{tabular}
	\begin{tabular}{ccccccccccc}
	$j$ & 2 & 3 & 4 & 5 & 6 & 7 & 8 & 9 & 10 \\
	\hline
	probability of $d_i=j$ & 10\% & 10\% & 10\% & 10\% & 10\% & 5\% & 5\% & 5\% & 35\% \\
	probability of $d_i\geq j$ & 100\% & 90\% & 80\% & 70\% & 60\% & 50\% & 45\% & 40\% & 35\% \\
\end{tabular}
	\caption{Probability mass function and proportions of individuals with at least $j$ measurements for the ``pruning distribution'' (top table: $d=5$; bottom table: $d=10$).}
	\label{tab:sim_study_freq}
\end{table}

For both the full and the pruned data set we use the sequential algorithm implemented in \texttt{RVineCopSelect} (from the \texttt{VineCopula} library) to fit D-vine copulas. In order to assess how badly the loss of information affects the estimation we compare the resulting D-vines by considering each pair-copula separately. For this purpose, we consider the mean absolute difference between the Kendall's $\tau$ values ($\Delta_\tau:=\frac{1}{R}\sum_{r=1}^R\left|\hat{\tau}_{\mathrm{pruned}}(r)-\hat{\tau}_{\mathrm{full}}(r) \right|$), the lower and the upper tail dependence coefficients ($\Delta_{\lambda^s}:=\frac{1}{R}\sum_{r=1}^R|\hat{\lambda}^s_{\mathrm{pruned}}(r)-\hat{\lambda}^s_{\mathrm{full}}(r)|$ for $s\in\left\{\ell,u\right\}$) of the two models. Comparing general strength of dependence and tail behavior enables us to assess how similar the fitted pair-copulas are.\footnote{Considering the percentage of cases where the same copula family is fitted would not be sensible since the number of candidate families is large and many of them, e.g.\ a Clayton and a survival Joe copula, can hardly be distinguished.} 

\paragraph{Results for $d=5$}\mbox{}\\
For $d=5$, the absolute differences of Kendall's $\tau$, lower and upper tail dependence coefficient (averaged over the $R=1000$ data sets) are displayed for each of the 10 pair-copulas in \autoref{tab:d5_N0200_tau_low_up}, where the sample sizes are $n=200$ and $n=2000$, respectively. For $n=200$, the 10 average absolute estimated Kendall's $\tau$ values for the full data set ($\frac{1}{R}\sum_{r=1}^R\left|\hat{\tau}_{\mathrm{full}}(r) \right|$) lie between 0.345 and 0.394; the 10 average estimated upper and lower tail dependence coefficients for the 10 pair-copulas are between 0.075 and 0.108 ($\frac{1}{R}\sum_{r=1}^R\hat{\lambda}^\ell_{\mathrm{full}}(r)$) and 0.080 and 0.107 ($\frac{1}{R}\sum_{r=1}^R\hat{\lambda}^u_{\mathrm{full}}(r)$), respectively. For $n=2000$, the three ranges are fairly similar: $[0.338,0.422]$, $[0.081,0.992]$ and $[0.083,0.107]$, respectively.
\setlength{\tabcolsep}{.35em}
\begin{table}[ht]
	\centering
	\begin{tabular}{llrrrrrrrrrr}
		\hline
		\rule[-1ex]{0pt}{2.5ex}
		& & $c_{1,2}$ & $c_{2,3}$ & $c_{3,4}$ & $c_{4,5}$ & $c_{1,3;2}$ & $c_{2,4;3}$ & $c_{3,5;4}$ & $c_{1,4;2,3}$ & $c_{2,5;3,4}$ & $c_{1,5;2,3,4}$\\ 
		\hline
		\rule{0pt}{2.5ex}
		\rule[-1ex]{0pt}{2.5ex}
		\parbox[t]{2mm}{\multirow{3}{*}{\rotatebox[origin=c]{90}{\small $n=200$}}} & $\Delta_\tau$ & 0.000 & 0.016 & 0.026 & 0.035 & 0.017 & 0.027 & 0.036 & 0.032 & 0.043 & 0.058\\ 
		\rule[-1ex]{0pt}{2.5ex}
		& $\Delta_{\lambda^\ell}$ & 0.000 & 0.018 & 0.035 & 0.050 & 0.024 & 0.039 & 0.053 & 0.044 & 0.066 & 0.065\\ 
		\rule[-1ex]{0pt}{2.5ex}
		& $\Delta_{\lambda^u}$ & 0.000 & 0.022 & 0.030 & 0.041 & 0.024 & 0.037 & 0.058 & 0.048 & 0.067 & 0.065\\ 
		\hline
		\rule{0pt}{2.5ex}
		\rule[-1ex]{0pt}{2.5ex}
		\parbox[t]{2mm}{\multirow{3}{*}{\rotatebox[origin=c]{90}{\small $n=2000$}}} & $\Delta_\tau$ & 0.000 & 0.005 & 0.007 & 0.010 & 0.005 & 0.008 & 0.010 & 0.008 & 0.011 & 0.015\\ 
		\rule[-1ex]{0pt}{2.5ex}
		& $\Delta_{\lambda^\ell}$ & 0.000 & 0.004 & 0.008 & 0.012 & 0.009 & 0.011 & 0.016 & 0.008 & 0.021 & 0.015\\ 
		\rule[-1ex]{0pt}{2.5ex}
		& $\Delta_{\lambda^u}$ & 0.000 & 0.004 & 0.010 & 0.011 & 0.006 & 0.008 & 0.013 & 0.011 & 0.021 & 0.024\\ 
		\hline
	\end{tabular}
	\caption{Absolute differences of Kendall's $\tau$, lower and upper tail dependence coefficient for each of the 10 pair-copulas, averaged over the $R=1000$ data sets of size $n=200$ and $n=2000$, respectively.}
	\label{tab:d5_N0200_tau_low_up}
\end{table}
\setlength{\tabcolsep}{6pt}

We can see that even for a sample size of only $n=200$ (see upper part of \autoref{tab:d5_N0200_tau_low_up}) the differences between the two estimates are relatively small. The largest absolute deviations are 0.058, 0.066 and 0.067 for $\tau$, $\lambda^\ell$ and $\lambda^u$, respectively. The average absolute deviations 0.029 ($\tau$), 0.039 ($\lambda^\ell$) and 0.039 ($\lambda^u$), respectively. Of course, $c_{1,2}$ is always estimated equally in both cases since all pruned observations have minimum length of 2. We can observe what one would expect given that the number of observations with at least $j$ measurements descends in $j$: Pair-copulas for whose estimation later measurements are needed exhibit larger deviations. 

The results in the lower part of \autoref{tab:d5_N0200_tau_low_up} (corresponding to $n=2000$) show a similar qualitative behavior. However, the overall level of average absolute deviations is even smaller: Maximum/average values are 0.015/0.008, 0.024/0.011 and 0.021/0.010 for $\tau$, $\lambda^\ell$ and $\lambda^u$, respectively.

\paragraph{Results for $d=10$}\mbox{}\\
We performed the same studies as above for $d=10$. Since it does not make sense to display the results for all 45 pair-copulas separately, we only report some summary statistics\footnote{The detailed results are of course available on request from the authors.}. For a sample size of $n=200$ the maximum/average deviations were 0.091/0.0483 ($\tau$), 0.069/0.044 ($\lambda^\ell$) and 0.068/0.043 ($\lambda^u$); for $n=2000$ we observed 0.059/0.018 ($\tau$), 0.037/0.017 ($\lambda^\ell$) and 0.037/0.017 ($\lambda^u$). In comparison to the results for $d=5$ we detect an increase in deviation, which seems plausible since the dimension of the model increases but the sample sizes are kept constant.\\

All in all, we see that the sequential fitting of D-vine models to repeated measurement data performs well such that we do not have to hesitate to use it for the real data application in \autoref{sec:application}.

\section{Application}\label{sec:application}
In \autoref{sec:simstudy} we have seen that our proposed estimation method performs satisfactory. Now we will apply it to real life data. For this purpose, we consider the aortic valve replacement surgery data set \texttt{heart.valve} that is taken from the R library \texttt{joineR} \citep{joineR} and has been analyzed in \cite{lim2008longitudinal}. For this longitudinal study the regression of the left ventricular mass index (LVMI) of $n=256$ individuals was examined in several follow-up appointments after the surgery, where a new heart valve had been implanted. The total number of examinations is 988 such that the average number of measurements per patient is 3.86, where 10 is the maximum. \autoref{tab:measurement_freq} summarizes the sizes of the groups of individuals with exactly $j$ and $j$ or more measurements, respectively, $j=1,\ldots,10$.\\
\begin{table}[h!]
	\centering
	\begin{tabular}{ccccccccccc}
	$j$ & 1 & 2 & 3 & 4 & 5 & 6 & 7 & 8 & 9 & 10 \\
	\hline
	patients with $j$ measurements & 51 & 44 & 41 & 30 & 27 & 15 & 21 & 15 & 6 & 6 \\
	patients with $\geq j$ measurements & 256 & 205 & 161 & 120 & 90 & 63 & 48 & 27 & 12 & 6 \\
	\end{tabular}
	\caption{Sizes of the groups of individuals with exactly $j$ and $j$ or more measurements, respectively, $j=1,\ldots,10$.}
	\label{tab:measurement_freq}
\end{table}

Besides the examination results, for every patient and measurement there are also covariates available. We denote them the way they are stored in the data set \texttt{heart.valve}. The following list contains the covariates that we used in our final models as well as a short description, which is basically taken from the documentation of the \texttt{joineR} library \citep{joineR}:
\begin{itemize}
	\item \texttt{size}: size of the heart valve in millimeters;
	\item \texttt{sex}: gender of the patient
	\item \texttt{bsa}: body surface area (preoperative)
	\item \texttt{time}: date of measurement (with surgery date as time origin)
\end{itemize}

The quantity we model is the logarithm of the LMVI. We estimate two different models: a linear mixed model and a D-vine copula based model. We focus on the first five measurements since there are rather few observations for the later measurements. This way we use 832 of the 988 available measurements ($84.2\%$). 

\paragraph{Linear mixed model approach}\mbox{}\\
In order to fit a linear mixed model (cf.\ \autoref{sec:LMM}) to the data we use the function \texttt{lme} from the R library \texttt{nlme} \citep{nlme}. Assuming a homogeneous covariance structure for all individuals $i$, $i=1,\ldots,256$, different correlation structures such as i.i.d.\ errors, compound symmetry or AR(1) can be selected (cf.\ \autoref{sec:connectionLMM}, \autopageref{list:corstructures}). We fit a random intercept model, i.e.\ $Z_i=(1,\ldots,1)^\top\in\Rbb^{d_i\times 1}$, and compare different (homogeneous) correlation structures for the error terms, namely i.i.d., compound symmetry, AR(1), exponential decay and general (i.e.\ unrestricted) structure. Note that we perform classical maximum-likelihood estimation (instead of restricted maximum-likelihood estimation, which is often used for linear mixed models) since we need be able to compare the quality of the fit to our D-vine based approach later. The parameter estimates, however, are almost the same for. The best model with respect to log-likelihood and AIC is the one with the general structure; it contains the covariates \texttt{size}, \texttt{sex} and \texttt{bsa} as well as an \texttt{intercept} as fixed effects. The AR(1) error structure, where the $(k,\ell)$th entry of $\Sigma_i$ is given by $\sigma^2\rho^{\left|k-\ell\right|}$, is more parsimonious than the general structure and exhibits a better BIC although log-likelihood and AIC are worse; it contains the covariates \texttt{size}, \texttt{sex}, \texttt{bsa} and \texttt{time} as well as an \texttt{intercept}. Note that the use of BIC for linear mixed models is controversial \citep{hedeker2006longitudinal}; it is frequently debated which sample size to use for the calculation of BIC \citep{jones2011bayesian,muller2013model,delattre2014note}. In the penalty of the standard BIC all model parameters are weighted with the logarithm of the total number of observations. Here, we used the adjusted BIC for linear mixed models that was developed by \cite{delattre2014note} and that is better comparable to the one we derived for our approach in \autoref{prop:BIC}, where each parameter is weighted with the logarithm of the number of observations that directly contribute to its estimation. In the adjusted penalty term of \cite{delattre2014note} the parameters associated with the fixed effects are weighted with the logarithm of the number of measurements and the parameters associated with the random effects are weighted with the logarithm of the number of individuals.  

The log-likelihood, AIC and BIC values of the models with the general and the AR(1) structure can be found in \autoref{tab:llAICBIC}. The remaining structures are not listed there as they performed uniformly worse than the two models. 

\begin{figure}[h!]
	\centering
	\includegraphics[width=\textwidth]{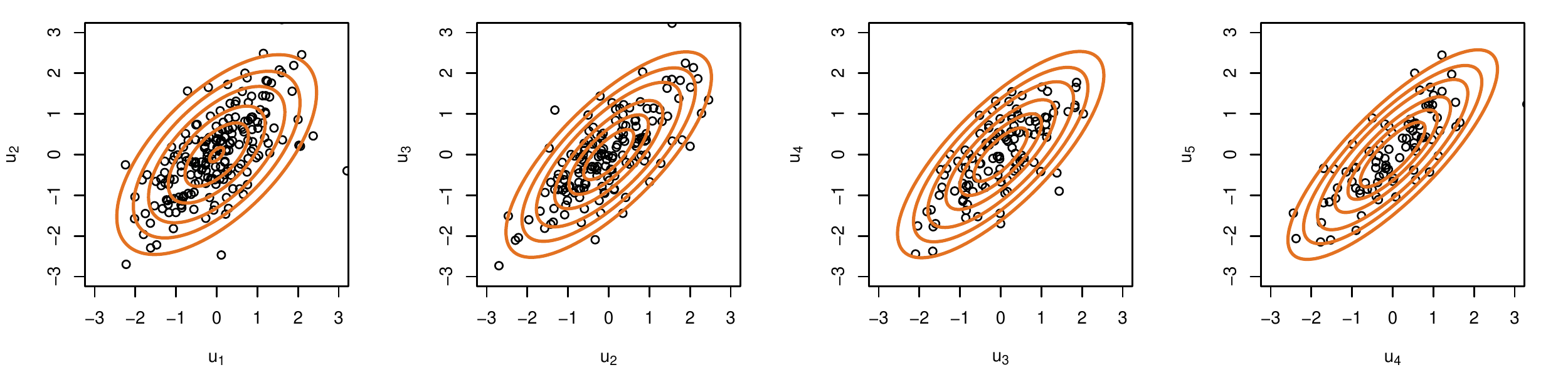}
	\includegraphics[width=\textwidth]{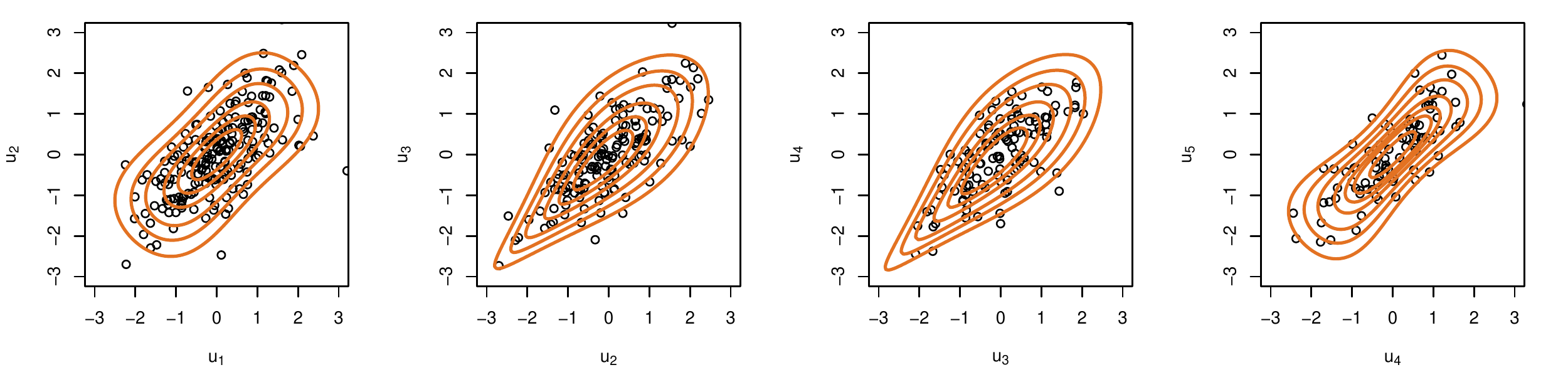}
	\caption{Pairwise plots of the copula data (transformed to standard normal margins) including the contours lines of the fitted pair-copulas of the Gaussian D-vine copula (upper panel) and the general D-vine copula (lower panel), respectively.}
	\label{fig:contours_pairs}
\end{figure}

\paragraph{D-vine copula based approach}\mbox{}\\
As an alternative we will also fit our D-vine based model. As described in \autoref{sec:setting} we first deal with the univariate marginal distributions and afterwards estimate the dependence structure. For the marginals we use the univariate marginal regression model that was already estimated for the linear mixed model with AR(1) error correlation structure. Hence, the margins depend on the covariates \texttt{size}, \texttt{sex}, \texttt{bsa} and \texttt{time}. In order to transform the measurements to the uniform scale, we apply the estimated normal distribution functions resulting from the regression model (cf.\ \autoref{eq:LMMdistribution}). 
Then a D-vine copula with order 1--2--3--4--5 is fitted to the transformed observations according to the sequential approach from \autoref{sec:depmodeling} (using \texttt{RVineCopSelect}). In order to avoid unnecessary parameters we apply a Kendall's $\tau$ based independence test (significance level $\alpha=5\%$), which is also implemented in \texttt{RVineCopSelect}, to decide for each pair-copula if it is significantly different from an independence copula \citep[for a detailed description see][]{hollander2014nonparametric,genest2007everything}. The criterion for the selection of the pair-copula families is standard BIC. We fit both a Gaussian D-vine copula, where all pairs are assumed to be bivariate Gaussian, and a general, unrestricted D-vine copula. The result is in both cases a first-order Markov structure, also known as a 1-truncated vine copula, i.e.\ all pair-copulas in the second, third and fourth tree are the independence. For the Gaussian vine the Kendall's $\tau$ values of the Gaussian pairs in the first tree are estimated to be $\hat{\tau}_{1,2}=0.43$, $\hat{\tau}_{2,3}=0.54$, $\hat{\tau}_{3,4}=0.56$ and $\hat{\tau}_{4,5}=0.61$. For the general D-vine copula the pair-copulas in the first tree are estimated to be the following: $\hat{c}_{1,2}=$ Frank ($\hat{\tau}_{1,2}=0.49$); $\hat{c}_{2,3}=$ Survival Gumbel ($\hat{\tau}_{2,3}=0.53$); $\hat{c}_{3,4}=$ Survival Gumbel ($\hat{\tau}_{3,4}=0.56$);  $\hat{c}_{4,5}=$ Frank ($\hat{\tau}_{4,5}=0.65$). \autoref{fig:contours_pairs} displays pairwise plots of the copula data (transformed to standard normal margins for reasons of comparability) including the contour lines of the corresponding fitted pair-copulas.

We see that there is a positive medium strength of dependence for all pairs in both models ($\tau$-values from 0.43 to 0.61 and 0.49 to 0.65, respectively). The shape of the contours, however, differs considerably between the two models. All four bivariate copulas in the general D-vine model are from different families and non-Gaussian. Whereas $\hat{c}_{1,2}$ and $\hat{c}_{4,5}$ show no tail dependence, the survival Gumbel copula modeling the dependence between the second and the third and the third and the fourth variable exhibits moderate lower tail dependence: $\hat{\lambda}_{2,3}^\ell=0.62$ and $\hat{\lambda}_{3,4}^\ell=0.64$. The fact that the dependence between two consecutive measurements is not constant and non-Gaussian is an indicator that the general D-vine approach might be a better choice than a simple Gaussian dependence model.\\

\paragraph{Model comparison}\mbox{}\\
In order to see if this is the case we compare the fit of the two D-vine based models (including margins) and the two linear mixed models to the data using the log-likelihood (of the full model including margins), AIC and BIC. \autoref{tab:llAICBIC} displays all three model selection criteria.
Note that the BIC values of the linear mixed models are calculated as proposed by \cite{delattre2014note} and the ones of the D-vine based models are calculated according to \autoref{prop:BIC}.
\begin{table}[h!]
	\centering
	\begin{tabular}{crrrr}
		model & log-likelihood & AIC & BIC & \# parameters \\
		\hline
		general LMM & $-99.4$ & $230.9$ & $305.3$ & $16$\\
		LMM AR(1) & $-108.9$ & $233.8$ & $270.4$ & $8$\\
		Gaussian D-vine & $-102.7$ & $225.3$ & $265.3$ & $10$\\
		general D-vine & $\mathbf{-85.0}$ & $\mathbf{190.1}$ & $\mathbf{230.1}$ & $10$\\
	\end{tabular}
	\caption{Log-likelihood, AIC and BIC values for the fitted linear mixed models with general and AR(1) structure and the Gaussian and the general D-vine based models (including margins). Bold values indicate the best model fit according to the respective model selection criteria.}
	\label{tab:llAICBIC}
\end{table}

One can see that the unrestricted D-vine model performs uniformly better than the Gaussian one. This is a clear indicator that the normality assumption for the dependence is not really suited. Nevertheless, due to their flexibility, both D-vine based models yield a considerably better fit than the two linear mixed model with respect to log-likelihood, AIC and BIC; only the log-likelihood of the Gaussian D-vine is slightly worse than the one of the general LMM. We see that the D-vine based approaches are able to capture the structure of the data better since the flexibility of the D-vine helps to fit the dependence structure more appropriately. This is especially important if the deviation from Gaussianity is strong. 

\paragraph{Quantile prediction}\mbox{}\\
As a final application we illustrate how conditional quantiles for the $(j+1)$st measurement of an individual with $j$ measurements can be determined using the general D-vine copula based approach and the linear mixed model with AR(1) error structure. For this purpose, we select three representative individuals with $d_i=4$ measurements from the data set such that they have had rather low ($\yb^1=(4.63, 4.62, 4.66, 4.91)^\top$), medium ($\yb^2=(5.26,5.13, 5.00,5.19)^\top$) and high ($\yb^3=(5.90, 5.80, 5.67, 5.31)^\top$) measurement values so far, respectively. We will use the corresponding observed covariate values of $\xb^i=(x_{i,1},x_{i,2},x_{i,3},x_{i,4})^\top=(\texttt{size}_i, \texttt{sex}_i, \texttt{bsa}_i, \texttt{time}_i)^\top$ that are given by $\xb^1=(29,0,1.93,3.15)^\top$, $\xb^2=(25,0,1.65,5.48)^\top$ and $\xb^3=(25, 0, 1.71,3.19)^\top$, respectively. 

We pretend that the three selected individuals have only had three measurements so far. Then we predict the median, i.e.\ the $50\%$ quantile, and a $90\%$ confidence interval, i.e.\ the $5\%$ and the $95\%$ quantile, of the fourth measurement based on the three measurements $y^i_1,y^i_2,y^i_3$ for both models and compare the results to the true value of the fourth measurement. 

For the linear mixed model we know that the joint distribution of the measurements of one individual is a multivariate normal distribution with mean and variance as in \autoref{eq:LMMvecdistribution}. Having estimated the corresponding parameters we can easily determine the conditional distribution of the fourth measurement given the first three measurements, which is given by a univariate normal distribution \citep[see for example][Section 2.6]{joe2014dependence}. The quantiles of univariate normal distributions are known such that we can easily compute the desired quantities. The estimated median for individual $1$ is 4.78, the estimated $90\%$ confidence interval is given by $(4.37,5.18)$. For individuals 2 and 3 the medians are 5.21 and 5.73 and the confidence intervals are $(4.81, 5.61)$ and $(5.32,6.13)$, respectively. We see that the observed value of the fourth measurement of individuals 1 and 2 are inside the confidence bounds; for the third individual the confidence interval does not contain the observed measurement value: $5.31\notin(5.32,6.13$). Note that due to the normality of the conditional distribution the confidence bounds are symmetric around the mean (which is also the median).

For comparison we determine the conditional quantiles using the general D-vine based model. Since we want to apply \autoref{eq:condquantile}, which uses the inverse conditional distribution function on the copula level $C_{4|1:3}^{-1}(\,\cdot\,|F^i_1(y^i_1),F^i_2(y^i_2),F^i_{3}(y^i_{3}))$ and the inverse of the marginal distribution function of the fourth measurement $(F^i_4(\,\cdot\,))^{-1}$, we first of all transform the measurements 1, 2 and 3 to the copula scale by $\hat{u}_j^i:=\hat{F}^i_j(y_j^i)$, where $\hat{F}^i_j$ are the marginal linear regression estimates obtained from the linear mixed model. For each individual $i=1,2,3$ and $\alpha \in \left\{0.05,0.50,0.95\right\}$ we use the estimated general D-vine copula to calculate $\hat q^i_u(\alpha):= \hat C_{4|1:3}^{-1}(\alpha|u^i_1,\ldots,u^i_{3})$. Since $\hat F^i_4(\,\cdot\,)$ is again a univariate normal distribution we can easily determine its inverse and apply $(\hat F^i_4(\alpha))^{-1}$ to $\hat q^i_u(\alpha)$ to obtain the conditional quantiles $\hat q^i_y(\alpha)=(\hat F^i_4(\alpha))^{-1}(\hat q^i_u(\alpha))$. The estimated median for individual $1$ is 4.65, the estimated $90\%$ confidence interval is given by $(4.42, 5.02)$. For individuals 2 and 3 the medians are 5.18 and 5.60 and the confidence intervals are $(4.82, 5.59)$ and $(5.13,6.03)$, respectively. Thus all three confidence intervals contain the corresponding observed measurement values. Note that the 5\% and 95\% quantiles are in general not symmetric around the 50\% quantile in this case.

\begin{figure}[h!]
	\centering
	\includegraphics[width=0.325\textwidth]{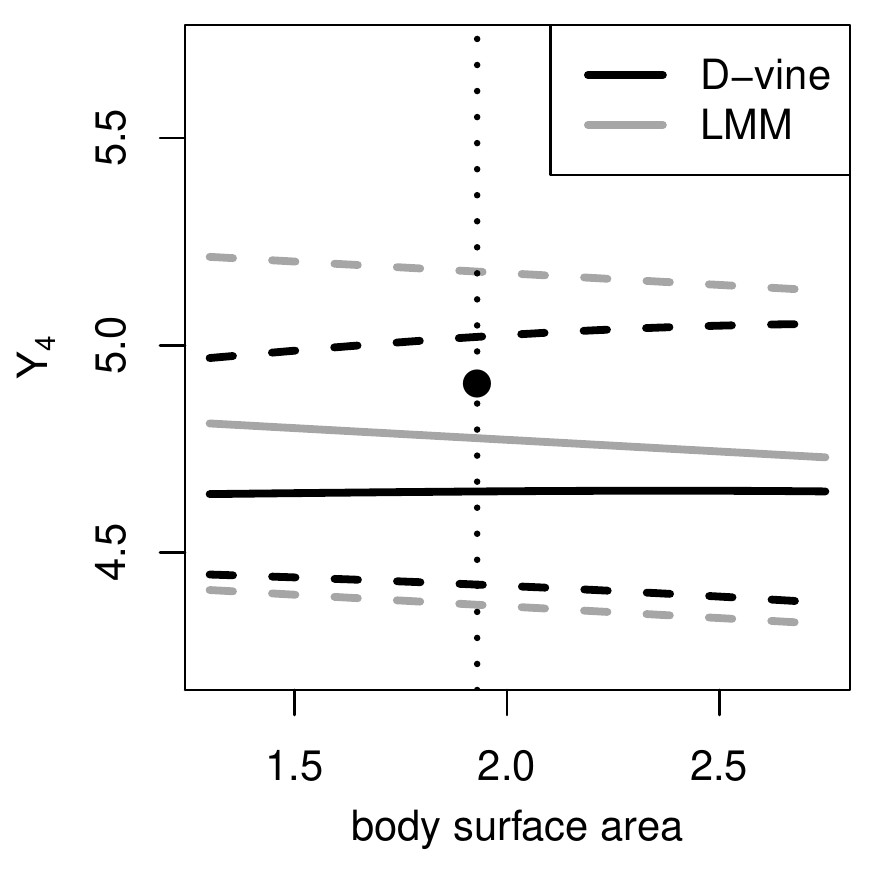}
	\includegraphics[width=0.325\textwidth]{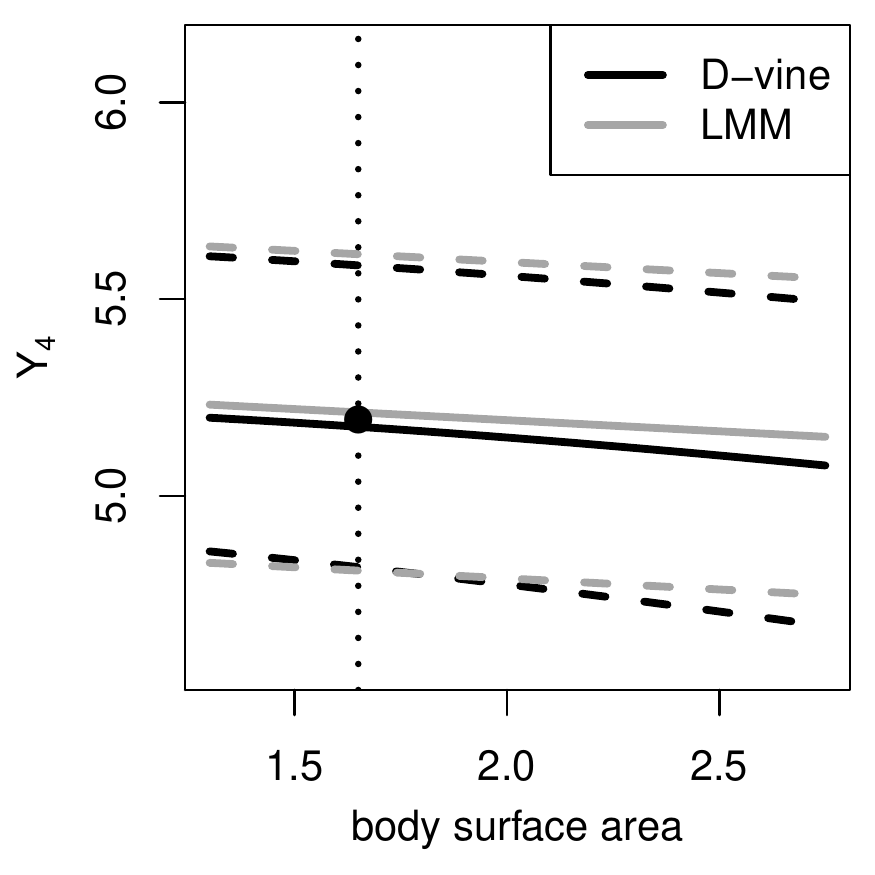}
	\includegraphics[width=0.325\textwidth]{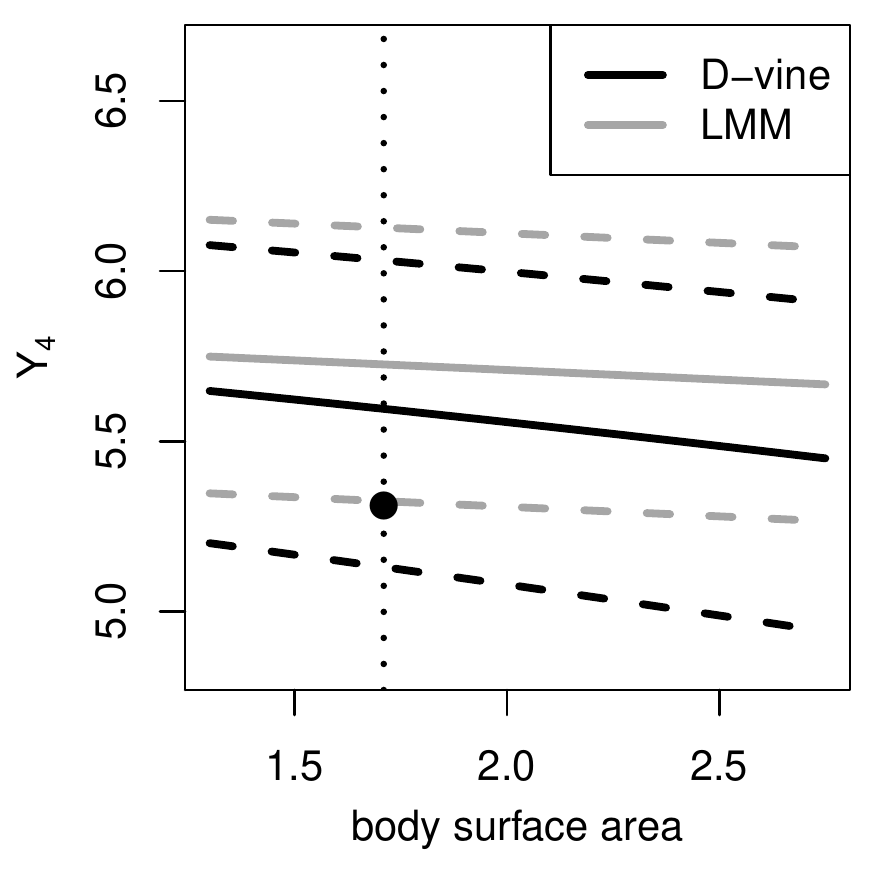}
	\caption{Estimated median (solid lines) and confidence intervals (dashed lines) of the fourth measurement for individuals 1 (left column), 2 (middle column) and 3 (right column) depending on the size of the heart valve for the LMM (gray) and the D-vine based model (black). Vertical lines indicate the true \texttt{bsa} values. Actual observed values are marked as circles.}
	\label{fig:prediction}
\end{figure}

In practice it might be interesting to investigate the influence of covariates on the estimated conditional quantiles. As an example we illustrate how the quantiles depend on the variable \texttt{bsa}. In \autoref{fig:prediction} we show the resulting estimates for the median (solid lines) and the confidence intervals (dashed lines) of individuals 1 (left), 2 (middle) and 3 (right) depending on the size of the heart valve for the linear mixed model (gray lines) and the D-vine based model (black lines). The estimated quantiles for the actual covariate specifications of our three selected individuals are given by the intersections dotted vertical lines (indicating the true \texttt{bsa} values) and the quantiles lines; for example the median is marked at the intersection point of the confidence interval and the corresponding median line. The actual observed measurement value $y^i_4$ is also added as a circle. 

First of all, we see that the quantiles estimated from the linear mixed model depend linearly on \texttt{bsa}. This is clear since the covariates only influence the estimated mean of the distribution of the fourth measurement given the first three. This influence is the same for all quantiles, i.e.\ the slope of the all gray lines is the same. Since the standard deviation does not depend on the covariate values all confidence intervals have the same width (even for different individuals). Normality implies that the confidence bounds lie symmetric around the mean (median).

The quantiles estimated on the basis of the D-vine based model, however, inherit the flexibility of the D-vine model and do not depend linearly on the \texttt{bsa} value. The width of the confidence intervals varies among the three individuals and even for one individual depending on the covariate values. The slope for different quantiles can even be positive and negative for one individual. These phenomena can be seen very clearly in left plot in \autoref{fig:prediction}) corresponding to the first individual.

This application illustrated how easily such investigations can be performed with both the linear mixed model and the D-vine based model. Analyzing the results, however, has made clear that the flexibility of the D-vine based model is a non-ignorable advantage over the linear mixed model. For both models it is eye-catching that the differences between the three individuals are considerable. This shows how important it can be to use the information available when making predictions for the future. 

\section{Conclusion and outlook}\label{sec:conclusion}
This article presented an intuitive and easily interpretable D-vine copula based model with arbitrary margins for (possibly) unbalanced longitudinal data. The model was compared to linear mixed models and proved to be a generalization of this model class under the assumption that the correlation structure was homogeneous over the individuals. Further, we developed a BIC adjustment for our model. Being based on D-vine copulas our proposed model benefited from the possibility to model the underlying dependence structure very flexibly. Since we did not impose any restrictions on the univariate marginal distributions, this adds even more flexibility to the model. As joint estimation of the D-vine copula would become rather slow in high dimensions, we proposed a fast sequential alternative, where even missing data values could be handled without causing problems. Due to the nested nature of D-vine models our approach further easily allowed for predicting future events. In the application to the heart surgery data set the proposed model was able to fit the data considerably better than the linear mixed models. If data exhibited an even more complicated dependence structure than the considered data set (possibly including stronger tail dependence, asymmetries etc.), the Gaussian assumption of linear mixed models would certainly be so strongly violated that changing to a more flexible model would be inevitable.

In an ongoing research project the D-vine based modeling approach is extended to time-to-event data with right-censoring \citep{barthel2017vine}.

\section*{Acknowledgments}
The first author acknowledges financial support by a research stipend of the Technical University of Munich. The second author is supported by the German Research Foundation (DFG Grant CZ 86/4-1). Numerical calculations were performed on a Linux cluster supported by DFG Grant INST 95/919-1 FUGG. The authors would like to thank Thomas Nagler for fruitful discussions and valuable comments on the paper.

\section*{Appendix}
\appendix

\section{Proof of \autoref{prop:BIC}}\label{App:proof_prop}
We will prove the statement of \autoref{prop:BIC} for $d=2$ in order to present the basic idea. The extension to higher dimensions works similarly but involves more tedious calculations. In our proof we adapt the derivation from \cite{neath2012bayesian}. Since our proof is very similar up to the last step, we refer the reader to their paper for a more detailed argumentation. \\

BIC is used for model selection when different parametric candidate models $M_1,\ldots,M_K$ are available to describe a data set $\Yb=\left\{\yb^1,\ldots,\yb^n\right\}$. 
Further, let $L(\thetab_k|\Yb)$ be the likelihood corresponding to model $M_k$, depending on the parameters $\thetab_k\in\Omega_k$, where $\Omega_k\subseteq\Rbb^{p_k}$ is the space of admissible parameters. Let $\pi(k)$ be the prior probability corresponding to model $M_k$ and $g(\thetab_k|k)$ denote a prior on $\thetab_k$ given the model $M_k$. Using Bayes' Theorem we obtain the joint posterior of $M_k$ and $\thetab_k$:
\[
h(k,\thetab_k|\Yc)=\frac{\pi(k)g(\thetab_k|k)L(\thetab_k|\Yc)}{m(\Yc)},
\]
where $m(\Yc)$ denotes the marginal distribution of $\Yb$. We are interested in finding the highest posterior probability of $M_k$ given $\Yc$, which can be expressed as
\[
P(k|\Yc)=\frac{\pi(k)}{m(\Yc)}\int_{\Omega_k}L(\thetab_k|\Yc)g(\thetab_k|k)\dint\thetab_k.
\]
Since maximizing $P(k|\Yc)$ is equivalent to minimizing $-2\log P(k|\Yc)$ with respect to $k$ and $m(\Yc)$ does not depend on $k$, we will from now on consider
\begin{equation}\label{eq:1}
S(k|\Yc):=-2 \log\pi(k)-2\log \int_{\Omega_k}L(\thetab_k|\Yc)g(\thetab_k|k)\dint\thetab_k.
\end{equation}
In order to be able to approximate the integrand from \autoref{eq:1} we perform a second-order Taylor series expansion of the log-likelihood $\log L(\thetab_k|\Yc)$ around the maximum-likelihood parameter estimate $\hat{\thetab}_k=\argmax_{\thetab_k\in\Omega_k}L(\thetab_k|\Yc)$:
\[
\begin{split}
\log L(\thetab_k|\Yc)\approx &\log L(\hat{\thetab}_k|\Yc)+(\thetab_k-\hat{\thetab}_k)^\top\frac{\partial \log L(\thetab_k|\Yc)}{\partial \thetab_k}\bigg|_{\thetab_k=\hat{\thetab}_k}\\
&+\frac{1}{2}(\thetab_k-\hat{\thetab}_k)^\top \left[ \frac{\partial^2\log L(\thetab_k|\Yc)}{\partial \thetab_k \partial \thetab_k^\top}\bigg|_{\thetab_k=\hat{\thetab}_k}\right](\thetab_k-\hat{\thetab}_k)
\end{split}
\]
Since $\hat{\thetab}_k$ maximizes $L(\thetab_k|\Yc)$, and hence also $\log L(\thetab_k|\Yc)$, we obtain 
\[
L(\thetab_k|\Yc)\approx L(\hat{\thetab}_k|\Yc)\exp\left\{-\frac{1}{2}(\thetab_k-\hat{\thetab}_k)^\top H(\hat{\thetab}_k|\Yc) (\thetab_k-\hat{\thetab}_k) \right\}
\]
where we denote the negative Hessian matrix of the log-likelihood by
\[
H(\thetab_k|\Yc):=-\frac{\partial^2\log L(\thetab_k|\Yc)}{\partial \thetab_k \partial \thetab_k^\top}.
\]
\cite{neath2012bayesian} and \cite{cavanaugh1999generalizing} argue that the above approximations hold for large samples $\Yc$ and further justify the use of a non-informative prior $g(\thetab_k|k)=1$ for any $\thetab_k\in\Omega_k$. Thus,
\begin{equation}\label{eq:2}
L(\thetab_k|\Yc)\approx L(\hat{\thetab}_k|\Yc) (2\pi)^{p_k/2}\left|H(\hat{\thetab}_k|\Yc)\right|^{-1/2}.
\end{equation}
Plugging \autoref{eq:2} into \autoref{eq:1} yields
\begin{equation}\label{eq:3}
S(k|\Yc)\approx -2\log\pi(k)-2\log L(\hat{\thetab}_k|\Yc) -p_k\log \pi +\log \left|H(\hat{\thetab}_k|\Yc)\right|.
\end{equation}
In order to compute the determinant of $H(\hat{\thetab}_k|\Yc)$ we consider the $(\ell,m)$th entry $H_{\ell,m}$ of $H(\thetab_k|\Yc)$. Since $d=2$ the parameter vector $\thetab_k=(\thetab_{k}^1,\thetab_{k}^2,\thetab_{k}^3)^\top$ can be split up such that $\thetab_{k}^j\in\Rbb^{q_j}$ parametrize the marginal distributions $F_j$ of the $j$th measurement, $j=1,2$ and $\thetab_{k}^3\in\Rbb^{q_3}$ is the parameter vector of the copula $c_{1,2}$ with $p_k=q_1+q_2+q_3$. For the sake of notation we assume that $\Yc$ is ordered such that $\Yc^2=\left\{\yb^1,\ldots,\yb^{n_2}\right\}$ and $\Yc^1=\left\{\yb^{n_2+1},\ldots,\yb^{n}\right\}$ and further recall that $N_1=n_1+n_2=n$ and $N_2=n_2$. We have
\[
\begin{split}
H_{\ell,m}&=-\frac{\partial^2}{\partial\theta_\ell\partial\theta_m}\sum_{i=1}^{n}\log L(\thetab_k|\yb^i)\\
&=-\sum_{i=1}^{N_1}\frac{\partial^2}{\partial\theta_\ell\partial\theta_m}\log f_1(y_1^i|\thetab_k^1) -\sum_{i=1}^{N_2}\frac{\partial^2}{\partial\theta_\ell\partial\theta_m}\log f_2(y_2^i|\thetab_k^2)\\&\phantom{=} -\sum_{i=1}^{N_2}\frac{\partial^2}{\partial\theta_\ell\partial\theta_m}\log c_{1,2}(F_1(y_1^i|\thetab_k^1), F_2(y_2^i|\thetab_k^2)|\thetab_k^3)\\
&=N_1\left[-\frac{1}{N_1}\sum_{i=1}^{N_1}\frac{\partial^2}{\partial\theta_\ell\partial\theta_m}\log f_1(y_1^i|\thetab_k^1) \right] +N_2\left[-\frac{1}{N_2}\sum_{i=1}^{N_2}\frac{\partial^2}{\partial\theta_\ell\partial\theta_m}\log f_2(y_2^i|\thetab_k^2) \right]\\
&\phantom{=}+N_2\left[-\frac{1}{N_2}\sum_{i=1}^{N_2}\frac{\partial^2}{\partial\theta_\ell\partial\theta_m}\log c_{1,2}(F_1(y_1^i|\thetab_k^1), F_2(y_2^i|\thetab_k^2)|\thetab_k^3) \right]\bigg.
\end{split}
\]
Assuming that the data set is large, i.e.\ $N_1$ and $N_2$ are large, the expressions in the brackets (approximately) represent entries of the Fisher information matrices 
\begin{align*}
I_1&=I_1(\thetab_k^1|\Yc)=-\Ebb\left[ \frac{\partial^2}{\partial\thetab^1_k\partial(\thetab^1_k)^\top}\log f_1(Y_1|\thetab_k^1) \right] \in \Rbb^{q_1\times q_1},\\
I_2&=I_2(\thetab_k^2|\Yc^2)=-\Ebb\left[ \frac{\partial^2}{\partial\thetab^2_k\partial(\thetab^2_k)^\top}\log f_2(Y_2|\thetab_k^2) \right] \in \Rbb^{q_2\times q_2}
\end{align*}
and 
\[
I_3=\begin{pmatrix}
I_3^{1,1} & I_3^{1,2} & I_3^{1,3} \\
I_3^{2,1} & I_3^{2,2} & I_3^{2,3} \\
I_3^{3,1} & I_3^{3,2} & I_3^{3,3}
\end{pmatrix}=I_3((\thetab_k^1,\thetab_k^2,\thetab_k^3)|\Yc^2)\in \Rbb^{(q_1+q_2+q_3)\times (q_1+q_2+q_3)},
\]
where 
\[
I_3^{\ell,m}=-\Ebb\left[ \frac{\partial^2}{\partial\thetab^\ell_k\partial(\thetab^m_k)^\top}\log c_{1,2}(F_1(Y_1|\thetab_k^1), F_2(Y_2|\thetab_k^2)|\thetab_k^3)  \right]\in \Rbb^{q_\ell\times q_m}.
\]
Thus, $H(\hat{\thetab}_k|\Yc)$ can be written as
\[
H(\hat{\thetab}_k|\Yc)=\begin{pmatrix}
N_1 I_1 + N_2 I_3^{1,1} & N_2 I_3^{1,2} & N_2 I_3^{1,3} \\
N_2 I_3^{2,1} & N_2 I_2 + N_2 I_3^{2,2} & N_2 I_3^{2,3} \\
N_2 I_3^{3,1} & N_2 I_3^{3,2} & N_2 I_3^{3,3}
\end{pmatrix}.
\]
Using the formula for the determinant of block-matrices \citep{silvester2000determinants} we obtain
\[
\begin{split}
\left|H(\hat{\thetab}_k|\Yc)\right|=&\,N_1^{q_1}N_2^{q_2+q_3}\bigg|\bigg.I_1+\frac{N_2}{N_1}I_3^{1,2}-I_3^{1,3}(I_3^{3,3})^{-1}I_3^{3,1}+\frac{N_2}{N_1}\left[I_3^{1,2}-I_3^{1,3}(I_3^{3,3})^{-1}I_3^{3,2} \right] \\
&\times \left[I_2+I_3^{2,2}-I_3^{2,3}(I_3^{3,3})^{-1}I_3^{3,2} \right]^{-1} \left[I_3^{2,1}-I_3^{2,3}(I_3^{3,3})^{-1}I_3^{3,1} \right]\bigg. \bigg| \\
&\times \Big| I_2+I_3^{2,2}-I_3^{2,3}(I_3^{3,3})^{-1}I_3^{3,2}\Big|\Big|I_3^{3,3}\Big|\\
&=:N_1^{q_1}N_2^{q_2+q_3}a(N_1,N_2).
\end{split}
\]
Note that since $N_2/N_1$ is bounded between 0 and 1, $a(N_1,N_2)$ is also bounded. Plugging the expression for $\left|H(\hat{\thetab}_k|\Yc)\right|$ into \autoref{eq:3} we obtain
\[
S(k|\Yc)\approx -2\log\pi(k)-2\log L(\hat{\thetab}_k|\Yc) -p_k\log \pi + q_1\log N_1 + (q_2+q_3)\log N_2 + \log a(N_1,N_2).
\]
Discarding the terms that are bounded as the sample size goes to infinity yields
\[
S(k|\Yc)\approx -2\log L(\hat{\thetab}_k|\Yc) + \Delta p_1\log N_1 + \Delta p_2\log N_2
\]
since $\Delta p_1=q_1$ and $\Delta p_2=q_2+q_3$. This proves the statement for $d=2$. The proof of \autoref{prop:BIC} in higher dimensions only differs from the above in that the calculations necessary to compute the determinant of $H(\hat{\thetab}_k|\Yc)$ are much more involved since one has to compute the determinant of a $(d(d+1)/2)\times (d(d+1)/2)$ block matrix.

\bibliographystyle{apalike}
\bibliography{References}

\end{document}